\documentclass[11pt,a4paper]{article}

\usepackage{jheppub}
\usepackage{amsmath,amssymb}
\usepackage[utf8]{inputenc}
\usepackage{graphicx}
\allowdisplaybreaks[1]

\title{Improved coarse-graining methods \\ on two dimensional tensor networks including fermions}

\author[a,b]{Muhammad Asaduzzaman,}
\author[a]{Simon Catterall,}
\author[b]{Yannick Meurice,}
\author[a]{Ryo Sakai,}
\author[a]{and Goksu Can Toga}

\affiliation[a]{Department of Physics, Syracuse University, Syracuse, NY 13244, USA}
\affiliation[b]{Department of Physics and Astronomy, The University of Iowa, Iowa City, Iowa 52242, USA}

\emailAdd{muhammad-asaduzzaman@uiowa.edu}
\emailAdd{smcatter@syr.edu}
\emailAdd{yannick-meurice@uiowa.edu}
\emailAdd{rsakai@syr.edu}
\emailAdd{gctoga@syr.edu}

\abstract{
  We show how to apply renormalization group algorithms incorporating entanglement filtering methods
  and a loop optimization to a tensor network which includes Grassmann variables which represent
  fermions in an underlying lattice field theory.
  As a numerical test a variety of quantities
  are calculated for two dimensional Wilson--Majorana fermions and for the two flavor
  Gross--Neveu model.  The improved algorithms show much better accuracy for
  quantities such as the free energy and the determination of Fisher's zeros.
}

\begin{document}
\maketitle

\section{Introduction}

The tensor renormalization group (TRG)~\cite{Levin:2006jai} has been very successfully applied to spin and gauge models in low dimensions~\footnote{
  The higher order TRG~\cite{2012PhRvB..86d5139X} can be used for higher dimensional systems, albeit numerical costs strongly depend on the dimensionality.
}.
It allows for high precision calculations of quantities including the free energy using very modest resources---see the recent review and references therein ref.~\cite{Meurice:2020pxc}.
The algorithm incorporates a parameter ---the bond dimension--- which controls convergence.
Close to a critical point, however, the required bond dimension grows large, and the algorithm becomes less efficient.
This has been traced to the fact that the standard TRG procedure can drive the coarsened tensor network to an artificial fixed point that is referred to as the corner double line (CDL) tensor~\cite{Gu:2009dr}~\footnote{
  The fixed point structure of the higher order TRG is also discussed in ref.~\cite{2014PhRvB..89g5116U}.
}.
To maintain a proper renormalization group flow improved algorithms have been developed, such as the tensor network renormalization (TNR)~\cite{2015PhRvL.115r0405E}, the loop-TNR~\cite{yang2017loop}, and the gilt-TNR~\cite{Hauru:2017tne} algorithm.
In such algorithms, entanglement filtering methods and further optimization steps are incorporated into the network coarsening procedure to eliminate the
CDL fixed point.

In this paper, we show how such algorithms can be adapted to the case of a tensor network involving Grassmann variables.
Grassmann tensor networks arise in generic lattice fermionic field theories and are needed to avoid sign problems.
In particular, in this paper, we test the combined Grassmann loop-TNR and Grassmann gilt-TNR algorithms in the case of two dimensional Wilson--Majorana fermions, which are known
to be equivalent to the two dimensional Ising model.

\section{TN representation for Wilson--Majorana fermions}

An equivalence between the Ising model and a lattice action for
Wilson--Majorana fermions has been shown for two dimensional honeycomb and square lattices~\cite{Wolff:2020oky}.
Here we construct a tensor network representation of the Wilson--Majorana
partition function on a square lattice.
While the equivalence has been shown in any choice of boundary conditions, throughout this paper, we assume the periodic and the antiperiodic boundary conditions for the spatial and the temporal directions, respectively.

The action of the Wilson--Majorana fermion model is given by
\begin{align}
  S =
  & \frac{1}{2} \sum_{n} \bar{\eta}_{n} \left( m_{\eta} + \sum_{\mu=1}^{2} \gamma_{\mu} \partial^{\mathrm{S}}_{\mu} - \frac{1}{2} \sum_{\mu=1}^{2} \partial_{\mu}\partial^{\ast}_{\mu} \right) \eta_{n}
    + \frac{1}{2} \sum_{n} \bar{\chi}_{n} \left( m_{\chi} + \sum_{\mu=1}^{2} \gamma_{\mu} \partial^{\mathrm{S}}_{\mu} - \frac{1}{2} \sum_{\mu=1}^{2} \partial_{\mu}\partial^{\ast}_{\mu} \right) \chi_{n} \nonumber \\
  & + \sum_{n} \bar{\eta}_{n} \left( \gamma_{1} \partial^{\mathrm{S}}_{1}
    - \gamma_{2} \partial^{\mathrm{S}}_{2}
    - \frac{1}{2} \partial_{1}\partial^{\ast}_{1}
    + \frac{1}{2} \partial_{2}\partial^{\ast}_{2} \right) \chi_{n}
\end{align}
with two component Majorana spinors $\eta \equiv (\eta_1,\eta_2)^{\rm{T}}$ and $\chi\equiv(\chi_1,\chi_2)^{\rm{T}}$.
The forward, the backward, and the symmetric difference operators are defined by $\partial$, $\partial^{\ast}$, and $\partial^{\mathrm{S}} = (\partial + \partial^{\ast})/2$, respectively.
The masses of the fermions are functions of the reverse temperature $\kappa$:
\begin{align}
  m_{\eta} = \frac{2}{\kappa}\left( \sqrt{2}-1-\kappa \right), && m_{\chi} = - \frac{2}{\kappa}\left( \sqrt{2}+1+\kappa \right).
\end{align}
The critical point of the system occurs for vanishing $m_{\eta}$, is given by $\kappa_{\mathrm{c}}=\sqrt{2}-1$, and corresponds to the critical point of the
two dimensional Ising model on a square lattice $\beta_{\mathrm{c}} = \tanh^{-1} \kappa_{\mathrm{c}}$.
Hereafter we use a representation of $\gamma$ matrices given by
\begin{align}
  \gamma_{1} = \sigma_{1} =
  \begin{pmatrix}
    0 & 1 \\
    1 & 0
  \end{pmatrix},
      && \gamma_{2} = \sigma_{3} =
         \begin{pmatrix}
           1 & 0 \\
           0 & -1
         \end{pmatrix},
      && C = -i\sigma_{2} =
         \begin{pmatrix}
           0 & -1 \\
           1 & 0
         \end{pmatrix}.
\end{align}
With this explicit representation, the Majorana condition turns out to be
\begin{align}
  \bar{\eta} = \eta^{\mathrm{T}}C = \left( \eta_{2} , -\eta_{1} \right)
\end{align}
together with the same condition on $\chi$.
Also, for simplicity,
we employ linear combinations of the spinor components. For example for the field $\eta$ we employ
\begin{align}
  \tilde{\eta} = \frac{1}{\sqrt{2}}
  \begin{pmatrix}
    \eta_{1}+\eta_{2} \\ -\eta_{1}+\eta_{2}
  \end{pmatrix}.
\end{align}
Then the Boltzmann factor can be expanded using integers that are placed on each link
\begin{align}
  e^{-S}
  = \sum_{\{x,t\}} \prod_{n}
  & e^{(m_{\eta}+2)\eta_{n,1}\eta_{n,2} + (m_{\chi}+2)\chi_{n,1}\chi_{n,2}}
    \left( -1 \right)^{x_{n,3}+x_{n,4}} \nonumber \\
  & \cdot \left( \tilde{\eta}_{n+\hat{1},2}\tilde{\eta}_{n,1} \right)^{x_{n,1}}
    \left( \tilde{\chi}_{n+\hat{1},2}\tilde{\chi}_{n,1} \right)^{x_{n,2}}
    \left( \tilde{\chi}_{n+\hat{1},2}\tilde{\eta}_{n,1} \right)^{x_{n,3}}
    \left( \tilde{\eta}_{n+\hat{1},2}\tilde{\chi}_{n,1} \right)^{x_{n,4}} \nonumber \\
  & \cdot \left( \eta_{n+\hat{2},2}\eta_{n,1} \right)^{t_{n,1}}
    \left( \chi_{n+\hat{2},2}\chi_{n,1} \right)^{t_{n,2}}
    \left( \chi_{n+\hat{2},2}\eta_{n,1} \right)^{t_{n,3}}
    \left( \eta_{n+\hat{2},2}\chi_{n,1} \right)^{t_{n,4}}.
\end{align}

To derive a tensor network representation we
introduce dummy Grassmann variables ($\Theta$ and $\bar{\Theta}$ in the following example) to split the hopping factors ($\Psi_{n+\hat{\mu}}\Phi_{n}$ in the following example) into products of commuting pairs by exploiting an identity
\begin{align}
  \Psi_{n+\hat{\mu}}\Phi_{n}
  = \int \left( \Psi_{n+\hat{\mu}}\mathrm{d}\bar{\Theta}_{n+{\mu}} \right)
  \left( \Phi_{n}\mathrm{d}\Theta_{n} \right)
  \left( \bar{\Theta}_{n+\hat{\mu}}\Theta_{n} \right).
\end{align}
In this way the partition function can be expressed as~\footnote{
  The expectation value of an observable can also be expressed as a tensor network in the presence of a so-called impurity tensor.
}
\begin{align}
  \label{eq:tnrep}
  Z &= \int \left( \prod_n \mathrm{d}\eta_{n,1}\mathrm{d}\eta_{n,2}\mathrm{d}\chi_{n,1}\mathrm{d}\chi_{n,2} \right) e^{-S} \nonumber \\
    &=  \sum_{\{x,t\}} \int \prod_{n} T_{x_{n} t_{n} x_{n-\hat{1}} t_{n-\hat{2}}} G_{n, x_{n} t_{n} x_{n-\hat{1}} t_{n-\hat{2}}},
\end{align}
where the bosonic tensor $T$, whose elements are numerical values, and the Grassmann valued tensor $G$ are defined by
\begin{align}
  \label{eq:deft}
  T_{ijkl} =
  \int \mathrm{d}\eta_{1}\mathrm{d}\eta_{2}\mathrm{d}\chi_{1}\mathrm{d}\chi_{2}
  & e^{(m_{\eta}+2)\eta_{1}\eta_{2} + (m_{\chi}+2)\chi_{1}\chi_{2}}
    \left( -1 \right)^{i_{3}+i_{4}} \nonumber \\
  & \cdot \eta_{2}^{l_{4}} \chi_{2}^{l_{3}} \chi_{2}^{l_{2}} \eta_{2}^{l_{1}}
    \tilde{\eta}_{2}^{k_{4}} \tilde{\chi}_{2}^{k_{3}} \tilde{\chi}_{2}^{k_{2}} \tilde{\eta}_{2}^{k_{1}}
    \chi_{1}^{j_{4}} \eta_{1}^{j_{3}} \chi_{1}^{j_{2}} \eta_{1}^{j_{1}}
    \tilde{\chi}_{1}^{i_{4}} \tilde{\eta}_{1}^{i_{3}} \tilde{\chi}_{1}^{i_{2}} \tilde{\eta}_{1}^{i_{1}},
\end{align}
and
\begin{align}
  \label{eq:defg}
  G_{n, ijkl} =
  & \mathrm{d}\alpha_{n,1}^{i_{1}} \cdots \mathrm{d}\alpha_{n,4}^{i_{4}}
    \mathrm{d}\beta_{n,1}^{j_{1}} \cdots \mathrm{d}\beta_{n,4}^{j_{4}}
    \mathrm{d}\bar{\alpha}_{n,1}^{k_{1}} \cdots \mathrm{d}\bar{\alpha}_{n,4}^{k_{4}}
    \mathrm{d}\bar{\beta}_{n,1}^{l_{1}} \cdots \mathrm{d}\bar{\beta}_{n,4}^{l_{4}} \nonumber \\
  & \cdot \left[ \prod_{s=1}^{4} \left( \bar{\alpha}_{n+\hat{1},s}\alpha_{n,s} \right)^{i_{s}} \left( \bar{\beta}_{n+\hat{2},s}\beta_{n,s} \right)^{j_{s}} \right].
\end{align}
Notice that the tensor indices have four components; \textit{e.g.} $i=(i_{1}, i_{2}, i_{3}, i_{4})$, where each component takes one of the
values $0$ or $1$ and expresses the fermion occupation.
Notice also that
$T$ is very sparse due to the nilpotency of the Grassmann variables in eq.~\eqref{eq:deft}. It is important to recognize that
all Grassmann variables are distinct, and \textit{e.g.} $\eta$ and $\bar{\eta}$ are not related by complex conjugation.
To contract this tensor network one will need to generalize the usual coarse-graining algorithms
to allow for blocking of both bosonic and Grassmann tensors.
While the ordering of the Grassmann variables matters in defining the tensor elements and in performing a coarse-graining procedure,
the final value of the contracted tensor network is invariant.

\section{Tensor renormalization group and entanglement filtering}

In this section we describe the usual TRG and explain its limitations
before reviewing the improved algorithms that circumvent these problems.
Initially we will focus on models without fermions before explaining
our current work to generalize these methods to Grassmann tensor networks.

\subsection{TRG and CDL structure}

Here we omit the Grassmann variables for simplicity
and assume that a general partition function $\mathcal{Z}$ is expressed as a tensor network
\begin{align}
  \mathcal{Z} = \sum_{\{x,t\}} \prod_{n} \mathcal{T}_{x_{n} t_{n} x_{n-\hat{1}} t_{n-\hat{2}}}.
\end{align}
The four rank tensor $\mathcal{T}$ can be decomposed by singular values in two ways:
\begin{align}
  & \mathcal{T}_{ijkl} = \sum_{m=1}^{D^{2}} U^{[1]}_{ijm} \sigma^{[13]}_{m} V^{[3] \dagger}_{mkl},
  & \mathcal{T}_{ijkl} = \sum_{m=1}^{D^{2}} U^{[2]}_{lim} \sigma^{[24]}_{m} V^{[4] \dagger}_{mjk},
\end{align}
where each index of $\mathcal{T}$ is assumed to run from $1$ to a certain bond dimension $D$
and where $\sigma$, $U$, $V$ are the singular values and unitary matrices arising
from a singular value decomposition (SVD).
By truncating the summations of the above decompositions according to
the magnitude of the singular values, the tensor
is approximately decomposed into two rank three tensors:
\begin{align}
  & \mathcal{T}_{ijkl} \approx \sum_{m=1}^{D} S^{[1]}_{ijm} S^{[3]}_{mkl},
  & \mathcal{T}_{ijkl} \approx \sum_{m=1}^{D} S^{[2]}_{lim} S^{[4]}_{mjk},
\end{align}
where $S^{[1]} = U^{[1]} \sqrt{\sigma^{[13]}}$, $S^{[2]} = U^{[2]} \sqrt{\sigma^{[24]}}$, $S^{[3]} = \sqrt{\sigma^{[13]}} V^{[3] \dagger}$, and $S^{[4]} = \sqrt{\sigma^{[24]}} V^{[4]}$.
Here the range of summation is truncated to $D$.
Of course the order of truncation is a tunable parameter, and the larger it is, the more accurate the approximation is.
It is known that the SVD minimizes the cost functions
\begin{align}
  & \left| \mathcal{T} - S^{[1]} S^{[3]} \right|_{\mathrm{F}},
  & \left| \mathcal{T} - S^{[2]} S^{[4]} \right|_{\mathrm{F}}
\end{align}
for a given truncation order~\cite{eckart1936approximation}.
Throughout this paper $\left| \ \cdot \ \right|_{\mathrm{F}}$ means the Frobenius norm.

By taking the SVDs for all tensors on the square lattice, the tensor network is approximately decomposed into a network of rank three tensors,
and then, by tracing out the old tensor indices, one can define a new rank four tensor
\begin{align}
  \mathcal{T}^{\star}_{ijkl} = \sum_{a,b,c,d=1}^{D} S^{[3]}_{icd} S^{[4]}_{jda} S^{[1]}_{abk} S^{[2]}_{bcl}
\end{align}
and obtain a coarse-grained square tensor network again.
Iterating this procedure many times
yields a small network, where one can exactly take the trace of all tensor indices and can obtain an approximate value of $\mathcal{Z}$.
This is the basic strategy of the TRG.

The SVD that is used above gives the optimal local
approximation for each tensor, but it is not optimal for the total tensor network.
Improved algorithms such as (loop-)TNR attempt to minimize a more global cost function as will be shown in the following section.

\begin{figure}[htbp]
  \centering
  \includegraphics[width=\hsize]{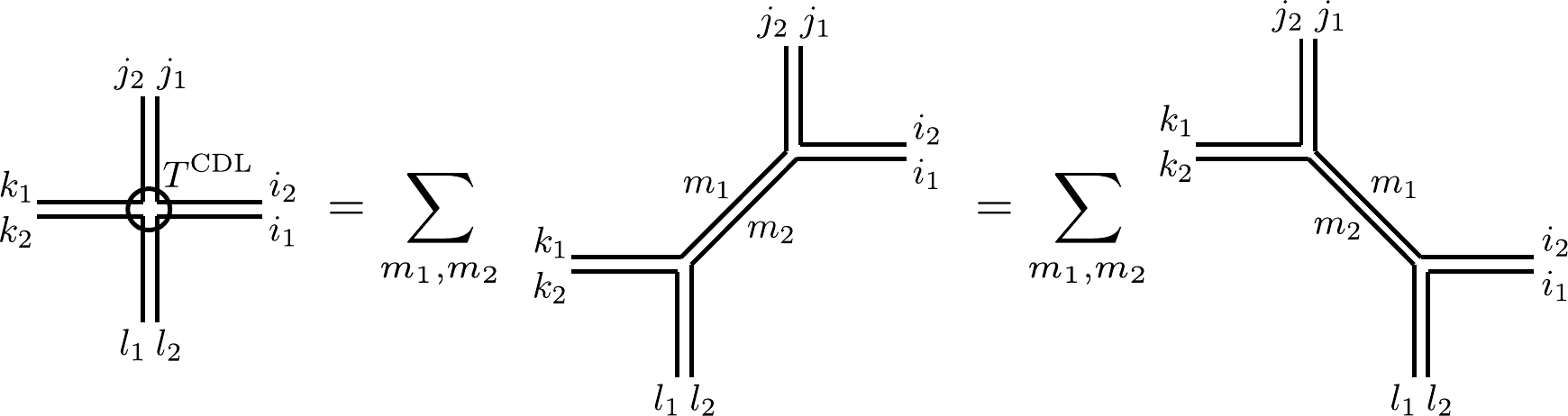}
  \caption{CDL tensor and its decompositions.}
  \label{fig:cdldecomposition}
\end{figure}

\begin{figure}[htbp]
  \centering
  \includegraphics[width=0.4\hsize]{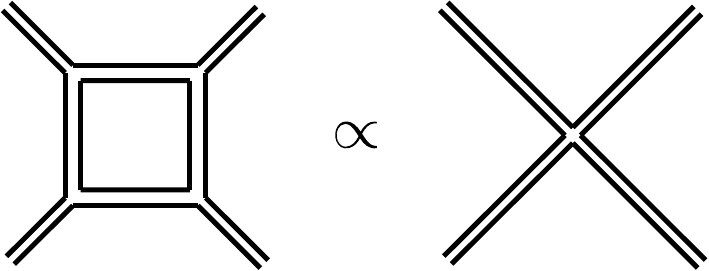}
  \caption{Coarse-graining of the CDL tensor network yields a CDL tensor network again.}
  \label{fig:coarsegrainedcdl}
\end{figure}

In addition to the problem of finding a better
cost function, the TRG has a known failure related to its fixed point structure.
As a typical example, one can easily show that a toy tensor network that consists of the CDL tensor
\begin{align}
  T^{\mathrm{CDL}}_{ijkl} = \Lambda_{i_{1}l_{2}} \Lambda_{j_{1}i_{2}} \Lambda_{k_{1}j_{2}} \Lambda_{l_{1}k_{2}},
\end{align}
where $\Lambda$ is a general matrix, which is often
simply set to be a diagonal matrix,
is a fixed point of the TRG~\cite{Gu:2009dr} (See figs.~\ref{fig:cdldecomposition}--\ref{fig:coarsegrainedcdl}.).
This fact means that the vanilla TRG leaves short range correlations in the network under a blocking step which, at a minimum, introduces a loss of accuracy into
the coarse-graining procedure and can even lead renormalization group flows into
artificial fixed points.
Any improved algorithm must have as its goal a procedure which eliminates this unphysical fixed point.

\subsection{Removal of CDL loops}

To remove the CDL structure from tensor networks, several improved coarse-graining methods have been developed so far.
In this section we review the loop-TNR~\cite{yang2017loop} and the gilt-TNR~\cite{Hauru:2017tne}.

\subsubsection{Loop-TNR (entanglement filtering and loop optimization)}

\begin{figure}[htbp]
  \centering
  \includegraphics[width=0.3\hsize]{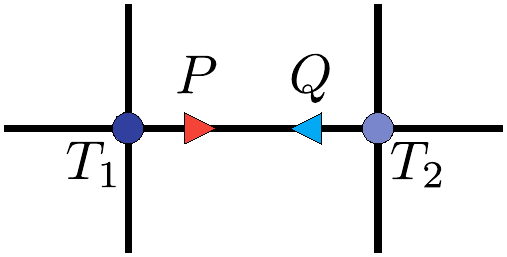}
  \caption{
    Insertion of projectors on a link.
    Here we assume that the tensor network has a checkerboard pattern, where $T_{1}$ and $T_{2}$ are placed at even sites and odd sites, respectively.
    The resulting tensor network after the loop optimization also has a checkerboard pattern.
  }
  \label{fig:projector}
\end{figure}

\begin{figure}[htbp]
  \centering
  \includegraphics[width=0.3\hsize]{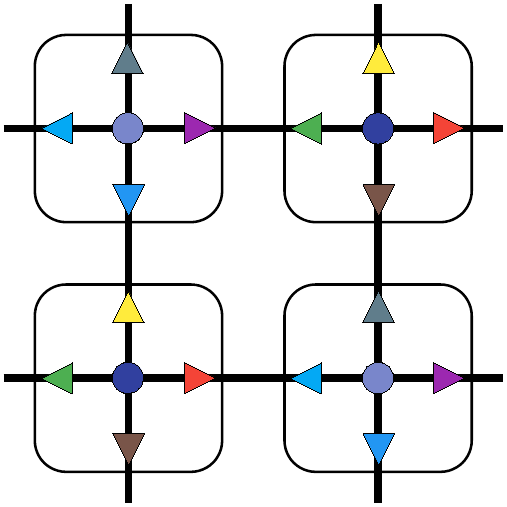}
  \caption{
    Gauge transformation of tensors on a plaquette.
    Tensors within lines are considered to be gauged tensors.
    This transformation removes the CDL loop from the plaquette.
  }
  \label{fig:gaugetrans}
\end{figure}

The loop-TNR algorithm
consists of an entanglement filtering step and a loop optimization step.
In the entanglement filtering step, projectors are inserted into each link (See figs.~\ref{fig:projector}--\ref{fig:gaugetrans}.).
Given a link the projectors are defined by
\begin{align}
  P = R Y \frac{1}{\sqrt{\tau}}, && Q = \frac{1}{\sqrt{\tau}} X^{\dag} L,
\end{align}
where the matrices $L$ and $R$ are constructed in an iterative procedure on a matrix product state with a periodic boundary condition and where $\tau$, $X$, and $Y$ are defined by the SVD of a matrix:
\begin{align}
  \left( LR \right) = X \tau Y.
\end{align}
The details of the construction of the matrices $L$ and $R$ are discussed in ref.~\cite{yang2017loop}.
In the reference there is also a proof of the fact that the given construction removes the CDL loop from a plaquette.
One can easily show that $P Q = L^{-1}LRR^{-1} = 1$;
thus the insertion of the projectors does not change the value of the partition function, and this step only takes place to ensure numerical stability.
In practice smaller values of $\tau$ are discarded, so that the SVD of $\left( LR \right)$ introduces a slight approximation with a truncation of the link; however, this step is considered to be just a gauge transformation of each tensor.

\begin{figure}[htbp]
  \centering
  \includegraphics[width=0.6\hsize]{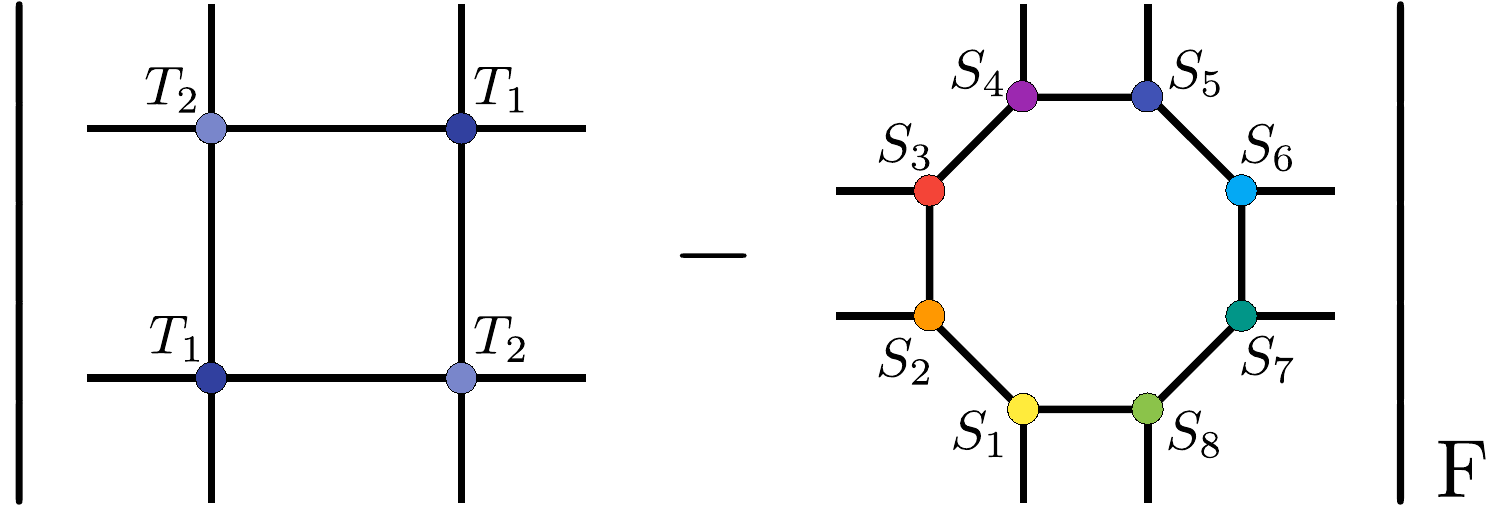}
  \caption{Cost function of the loop optimization step.}
  \label{fig:looptnrcost}
\end{figure}

In the loop optimization step, a set of rank three tensors $S_{1}$, $S_{2}$, $S_{3}$, $S_{4}$, $S_{5}$, $S_{6}$, $S_{7}$, and $S_{8}$ is tuned to approximate the original plaquette.
Namely, the cost function
\begin{align}
  \left| T_{1} T_{2} T_{1} T_{2} - S_{1} S_{2} S_{3} S_{4} S_{5} S_{6} S_{7} S_{8} \right|_{\mathrm{F}}
\end{align}
is minimized (See fig.~\ref{fig:looptnrcost}).
In this step the set $\left\{ S \right\}$ is minimized one by one while keeping all the other tensors fixed.
Under this situation, the tensor that is of interest is optimized through solving a set of simultaneous linear equations.

As for the initial condition of $\left\{ S \right\}$, it can be chosen from the truncated SVDs of the original rank four tensors as in the case of the usual TRG.
Indeed, as emphasized in the original paper~\cite{yang2017loop}, the choice of the initial condition affects the convergence behavior of the loop optimization step.
In ref.~\cite{Hong_2022}, where the classical XY model is analyzed, the octagonal tensor ring is prepared by the full SVDs without any approximation, and the bond dimensions are truncated by entanglement filtering projectors after that;
it has numerically shown that this choice improves the speed of convergence significantly.
In the numerical works in our current paper, we adopt their choice for the initial condition of the loop optimization step.

In practice, one can easily find convergence up to the machine precision in making the entanglement filtering projectors.
By contrast, it is hard to make the loop optimization step fully converge.
Then, in an actual implementation, one will set a maximum number of iterations or a sufficient condition for the cost function.
Of course these properties depend on the model and physical parameters.
In our current work, we quit the loop optimization step if the residual difference of the cost function gets as small as $10^{-4}$.

\subsubsection{Gilt-TNR (graph independent local truncations)}

The entanglement filtering scheme is not unique, and indeed
several algorithms have been proposed.
An example that may be graphically easy to understand is called the graph independent local truncation (gilt), which was proposed in ref.~\cite{Hauru:2017tne}.

\begin{figure}[htbp]
  \centering
  \includegraphics[width=0.5\hsize]{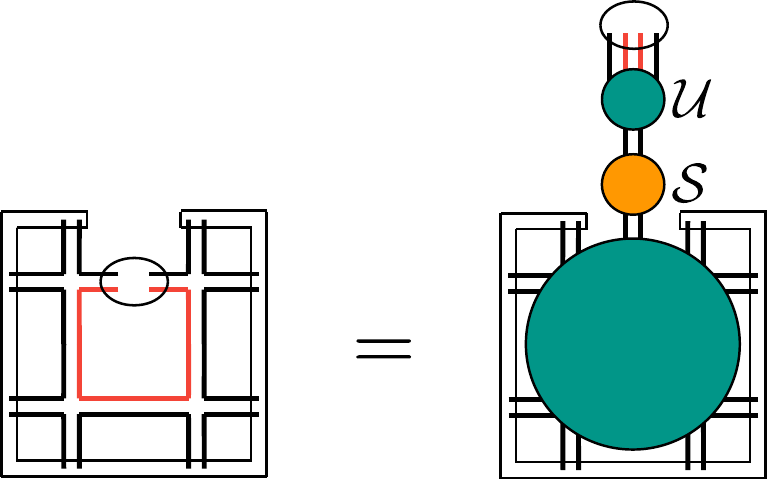}
  \caption{SVD of the environment of an open link on a plaquette. CDL loop is captured by the unitary matrix $\mathcal{U}$.}
  \label{fig:svdenv}
\end{figure}

In this method one considers a plaquette as an environment of a link.
By taking the SVD of the environment, the CDL loop is captured by a unitary matrix $\mathcal{U}$ (See fig.~\ref{fig:svdenv}).
To remove the CDL loop from the plaquette, $\mathcal{U}$ is replaced by another matrix according to a certain criterion with a hyperparameter $\epsilon_{\mathrm{gilt}}$.
In their method, a partial trace of $\mathcal{U}_{i}$ that is associated with $i$-th environment spectrum $\mathcal{S}_{i}$ is defined by
\begin{align}
  t_{i} = \mathrm{tr} \,\mathcal{U}_{i}
\end{align}
and is replaced by
\begin{align}
  t_{i}^{\prime} = t_{i} \frac{\mathcal{S}_{i}^{2}}{\mathcal{S}_{i}^{2} + \epsilon_{\mathrm{gilt}}^{2}}.
\end{align}
After this replacement, $t^{\prime}$ is pushed back to the link,
and this procedure is repeated until a convergence condition is satisfied.
The parameter
$\epsilon_{\mathrm{gilt}}$ determines how drastically the bond dimension of the link is truncated.
A detailed discussion is given in refs.~\cite{Hauru:2017tne,Delcamp:2020hzo}.
In particular, ref.~\cite{Delcamp:2020hzo} gives a hint of how best to tune $\epsilon_{\mathrm{gilt}}$.
In the following section we show numerical results with some representational choices of $\epsilon_{\mathrm{gilt}}$.

\subsection{Coarse-graining methods for fermion systems}

In fermion systems the tensor network factorizes into the product
of a bosonic and a Grassmann valued tensor
as discussed earlier.
To coarse grain such networks, one has to take care of the Grassmann variables.
Conceptually there are two ways to deal with Grassmann variables;
one is to coarse grain the Grassmann part separately from the bosonic tensors,
while the other introduces the Grassmann algebra directly into the tensor operations such as transpositions, multiplications, and decompositions~\cite{Gu:2010yh,Gu:2013gba}.
An important point is that the treatment of the Grassmann variables can be done without any approximation,
so that one yields the same numerical results whichever way is chosen.
In this paper we discuss the former approach.

\begin{figure}[htbp]
  \centering
  \includegraphics[width=0.8\hsize]{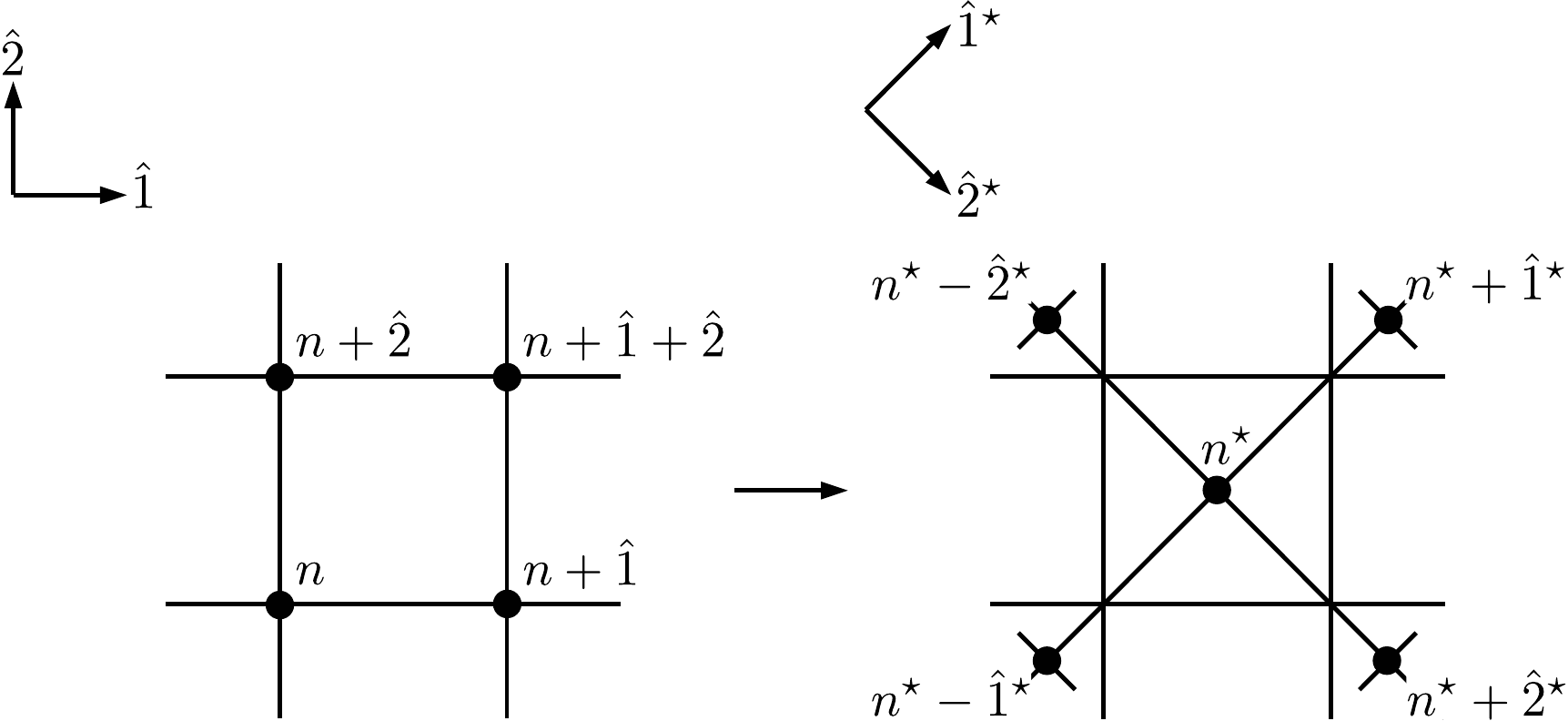}
  \caption{Coarse-grained lattice.}
  \label{fig:cglattice}
\end{figure}

The Grassmann part can be coarse grained in a way that goes along with the TRG.
First, $G$ is decomposed as
\begin{align}
  G_{n,ijkl} = \int \left( \Theta^{[1]}_{n,ij} \mathrm{d}\bar{\alpha}_{n^{\star}}^{m_{\mathrm{f}}} \right)
  \left( \Theta^{[3]}_{n,kl} \mathrm{d}\alpha_{n^{\star}-\hat{1}^{\star}}^{m_{\mathrm{f}}} \right)
  \left( \bar{\alpha}_{n^{\star}} \alpha_{n^{\star}-\hat{1}^{\star}} \right)^{m_{\mathrm{f}}},
\end{align}
where
\begin{align}
  \Theta^{[1]}_{n,ij}
  & = \mathrm{d}\alpha_{n,1}^{i_{1}} \cdots \mathrm{d}\alpha_{n,4}^{i_{4}}
    \mathrm{d}\beta_{n,1}^{j_{1}} \cdots \mathrm{d}\beta_{n,4}^{j_{4}}
    \left[ \prod_{s=1}^{4} \left( \bar{\alpha}_{n+\hat{1},s}\alpha_{n,s} \right)^{i_{s}} \left( \bar{\beta}_{n+\hat{2},s}\beta_{n,s} \right)^{j_{s}} \right], \\
  \Theta^{[3]}_{n,kl}
  & = \mathrm{d}\bar{\alpha}_{n,1}^{k_{1}} \cdots \mathrm{d}\bar{\alpha}_{n,4}^{k_{4}}
    \mathrm{d}\bar{\beta}_{n,1}^{l_{1}} \cdots \mathrm{d}\bar{\beta}_{n,4}^{l_{4}}
\end{align}
with
\begin{align}
  m_{\mathrm{f}} = \left( \sum_{s=1}^{4} i_{s} + \sum_{s=1}^{4} j_{s} \right) \bmod 2 = \left( \sum_{s=1}^{4} k_{s} + \sum_{s=1}^{4} l_{s} \right) \bmod 2.
\end{align}
The new symbol $n^{\star}$ denotes a site on the coarse-grained lattice,
and the unit vectors on the coarse-grained lattice are defined by $\hat{1}^{\star} = \hat{1} + \hat{2}$ and $\hat{2}^{\star} = \hat{2} - \hat{1}$ here~\footnote{
  Of course the definition of new unit vectors are one's option.
  Another option would be a (anti)clockwise orientation.
}.
The graphical definition of the coarse-grained lattice is given in fig.~\ref{fig:cglattice}.
$\alpha_{n^{\star}}$, $\bar{\alpha}_{n^{\star}}$, $\beta_{n^{\star}}$, and $\bar{\beta}_{n^{\star}}$ are dummy Grassmann variables that are placed on the coarse-grained network.

$G$ is decomposed in a similar way as
\begin{align}
  G_{n,ijkl} = \left( -1 \right)^{\sum_{s=1}^{4} l_{s}} \int \left( \Theta^{[2]}_{n,li} \mathrm{d}\bar{\beta}_{n^{\star}}^{m_{\mathrm{f}}} \right) \left( \Theta^{[4]}_{n,jk} \mathrm{d}\beta_{n^{\star}-\hat{2}^{\star}}^{m_{\mathrm{f}}} \right) \left( \bar{\beta}_{n^{\star}} \beta_{n^{\star}-\hat{2}^{\star}} \right)^{m_{\mathrm{f}}},
\end{align}
where
\begin{align}
  & \Theta^{[2]}_{n,li}
    = \mathrm{d}\bar{\beta}_{n,1}^{l_{1}} \cdots \mathrm{d}\bar{\beta}_{n,4}^{l_{4}}
    \mathrm{d}\alpha_{n,1}^{i_{1}} \cdots \mathrm{d}\alpha_{n,4}^{i_{4}}
    \left[ \prod_{s=1}^{4} \left( \bar{\alpha}_{n+\hat{1},s}\alpha_{n,s} \right)^{i_{s}} \right], \\
  & \Theta^{[4]}_{n,jk}
    = \mathrm{d}\beta_{n,1}^{j_{1}} \cdots \mathrm{d}\beta_{n,4}^{j_{4}}
    \mathrm{d}\bar{\alpha}_{n,1}^{k_{1}} \cdots \mathrm{d}\bar{\alpha}_{n,4}^{k_{4}}
    \left[ \prod_{s=1}^{4} \left( \bar{\beta}_{n+\hat{2},s}\beta_{n,s} \right)^{j_{s}} \right]
\end{align}
with
\begin{align}
  m_{\mathrm{f}} = \left( \sum_{s=1}^{4} l_{s} + \sum_{s=1}^{4} i_{s} \right) \bmod 2
  = \left( \sum_{s=1}^{4} j_{s} + \sum_{s=1}^{4} k_{s} \right) \bmod 2.
\end{align}

Using the components above one can define a coarse-grained Grassmann tensor $G^{\star}$:
\begin{align}
  &
    \begin{aligned}[t]
      & \left( \Theta^{[3]}_{n+\hat{1}+\hat{2},cd} \mathrm{d}\alpha_{n^{\star}}^{x_{n^{\star},\mathrm{f}}} \right)
      \left( \Theta^{[4]}_{n+\hat{1},da} \mathrm{d}\beta_{n^{\star}}^{t_{n^{\star},\mathrm{f}}} \right)
      \left( \Theta^{[1]}_{n,ab} \mathrm{d}\bar{\alpha}_{n^{\star}}^{x_{n^{\star}-\hat{1}^{\star},\mathrm{f}}} \right)
      \left( \Theta^{[2]}_{n+\hat{2},bc} \mathrm{d}\bar{\beta}_{n^{\star}}^{t_{n^{\star}-\hat{2}^{\star},\mathrm{f}}} \right) \\
      & \cdot \left( \bar{\alpha}_{n^{\star}+\hat{1}^{\star}} \alpha_{n^{\star}} \right)^{x_{n^{\star},\mathrm{f}}}
      \left( \bar{\beta}_{n^{\star}+\hat{2}^{\star}} \beta_{n^{\star}} \right)^{t_{n^{\star},\mathrm{f}}}
    \end{aligned}
  \\
  = & \Theta^{[2]}_{n+\hat{2},bc} \Theta^{[1]}_{n,ab} \Theta^{[4]}_{n+\hat{1},da} \Theta^{[3]}_{n+\hat{1}+\hat{2},cd}
      G^{\star}_{n,x_{n^{\star}} t_{n^{\star}} x_{n^{\star}-\hat{1}^{\star}} t_{n^{\star}-\hat{2}^{\star}}},
\end{align}
where $\Theta^{[2]}\Theta^{[1]}\Theta^{[4]}\Theta^{[3]}$ turns out to be a phase factor by integrating out the old Grassmann variables
and where~\footnote{
  Note that the coarse-grained tensor has less content than the original one.
  This shape is kept after this coarse-graining step.
}
\begin{align}
  G^{\star}_{n,x_{n^{\star}} t_{n^{\star}} x_{n^{\star}-\hat{1}^{\star}} t_{n^{\star}-\hat{2}^{\star}}}
  = \mathrm{d}\alpha_{n^{\star}}^{x_{n^{\star},\mathrm{f}}}
  \mathrm{d}\beta_{n^{\star}}^{t_{n^{\star},\mathrm{f}}}
  \mathrm{d}\bar{\alpha}_{n^{\star}}^{x_{n^{\star}-\hat{1}^{\star},\mathrm{f}}}
  \mathrm{d}\bar{\beta}_{n^{\star}}^{t_{n^{\star}-\hat{2}^{\star},\mathrm{f}}}
  \left( \bar{\alpha}_{n^{\star}+\hat{1}^{\star}} \alpha_{n^{\star}} \right)^{x_{n^{\star},\mathrm{f}}}
  \left( \bar{\beta}_{n^{\star}+\hat{2}^{\star}} \beta_{n^{\star}} \right)^{t_{n^{\star},\mathrm{f}}}.
\end{align}
Note that there is no approximation in the treatment of the Grassmann part,
and, after this step, one just obtains exact phase factors and constraints act on the bosonic tensors.
The phase factors generated from the Grassmann integrals are incorporated into the bosonic
tensor $T$, and then the coarse-graining step of $T$s takes place as usual via a suitable algorithm. Notice that
improved algorithms such as the loop-TNR and the gilt-TNR can be applied to the bosonic part of the tensor network at this stage.

\section{Numerical results}

\begin{figure}[htbp]
  \centering
  \includegraphics[width=0.8\hsize]{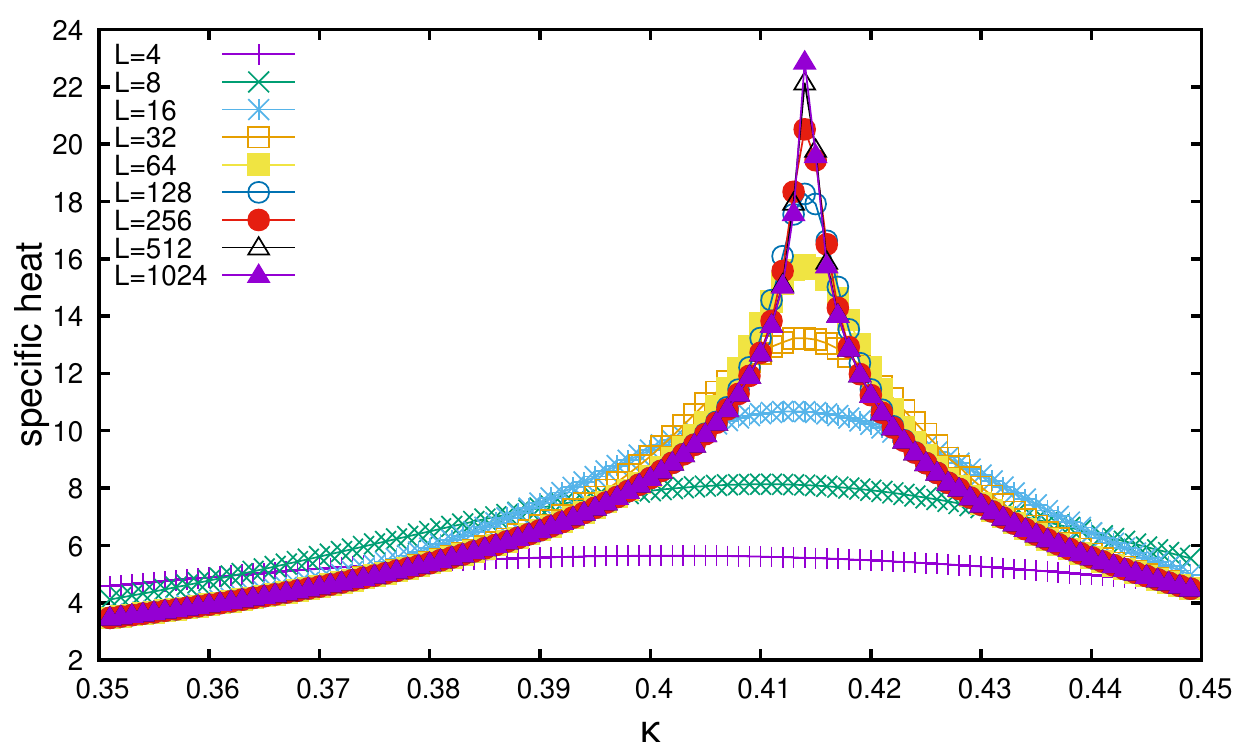}
  \caption{Specific heats of the Wilson--Majorana fermion system.}
  \label{fig:specheat_2dwilsonmajorana_bdim32}
\end{figure}

To demonstrate the equivalence to the Ising model, fig.~\ref{fig:specheat_2dwilsonmajorana_bdim32} shows the specific heat of the Wilson--Majorana fermion system.
Here the specific heat is calculated by taking the second numerical derivative of the free energy.
One can clearly observe a logarithmic growth of the peak height of the specific heat at the critical point $\kappa_{\mathrm{c}} = \sqrt{2}-1$.
The regions $\kappa < \kappa_{\mathrm{c}}$ and $\kappa > \kappa_{\mathrm{c}}$ correspond to the Z$_{2}$ symmetric (high temperature) and the broken (low temperature) phases, respectively.

\begin{figure}[htbp]
  \centering
  \includegraphics[width=0.8\hsize]{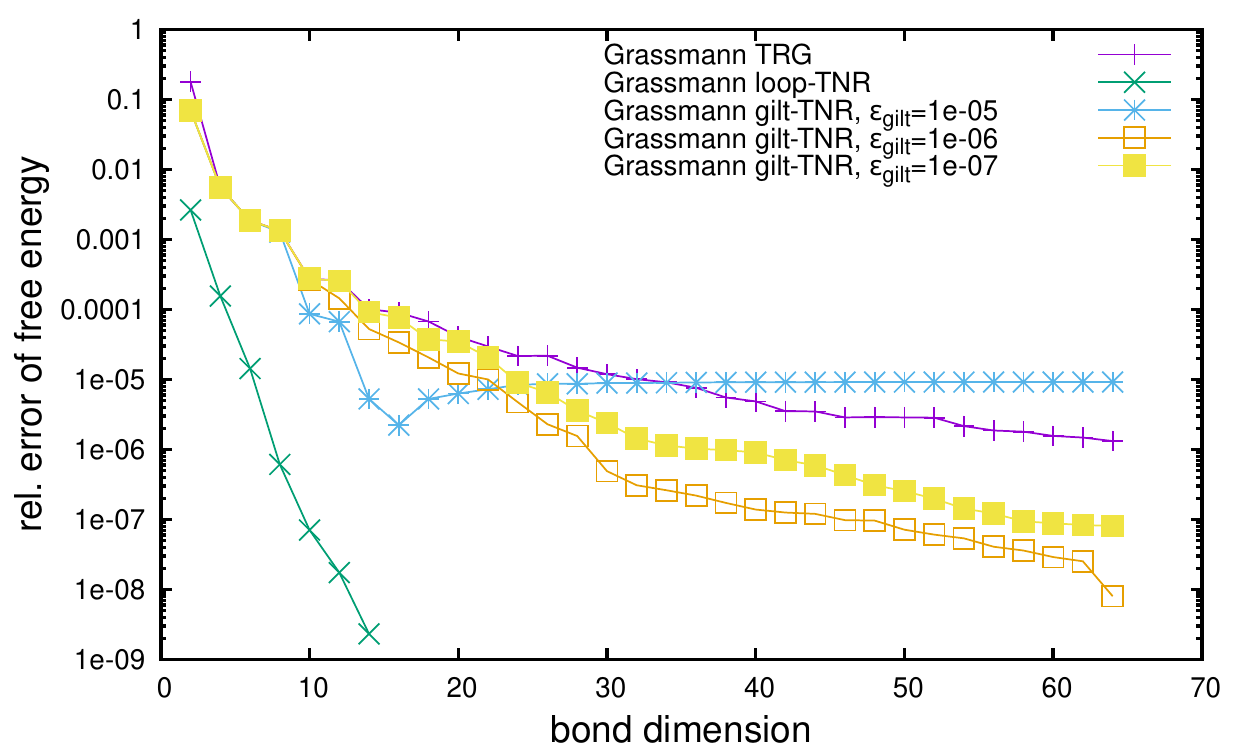}
  \caption{Relative errors of the free energy at the criticality. $L=1024$}
  \label{figs:relerrfeneg_2dwilsonmajorana_kappa0.4142135624_iter9}
\end{figure}

Figure~\ref{figs:relerrfeneg_2dwilsonmajorana_kappa0.4142135624_iter9} shows the relative errors of the free energy obtained by the plain Grassmann TRG and the improved ones.
The linear system size is set to $L=1024$ for this comparison.
Clearly the Grassmann loop-TNR result shows
an exponential improvement of accuracy as compared to the plain Grassmann TRG,
and the accuracy reaches better than single precision at a relatively small bond dimension like $10$.
The Grassmann gilt-TNR shows a similar accuracy to the vanilla Grassmann TRG at lower bond dimensions.
At larger bond dimensions, the accuracy of the Grassmann gilt-TNR depends on the hyperparameter $\epsilon_{\mathrm{gilt}}$.
Optimal $\epsilon_{\mathrm{gilt}}$ depends on the model, physical parameters, and even on the bond dimension.
In our experiments, $\epsilon_{\mathrm{gilt}} = 10^{-6}$ seems like a good choice near the
criticality in this model for large bond dimensions.

\begin{figure}[htbp]
  \centering
  \includegraphics[width=0.8\hsize]{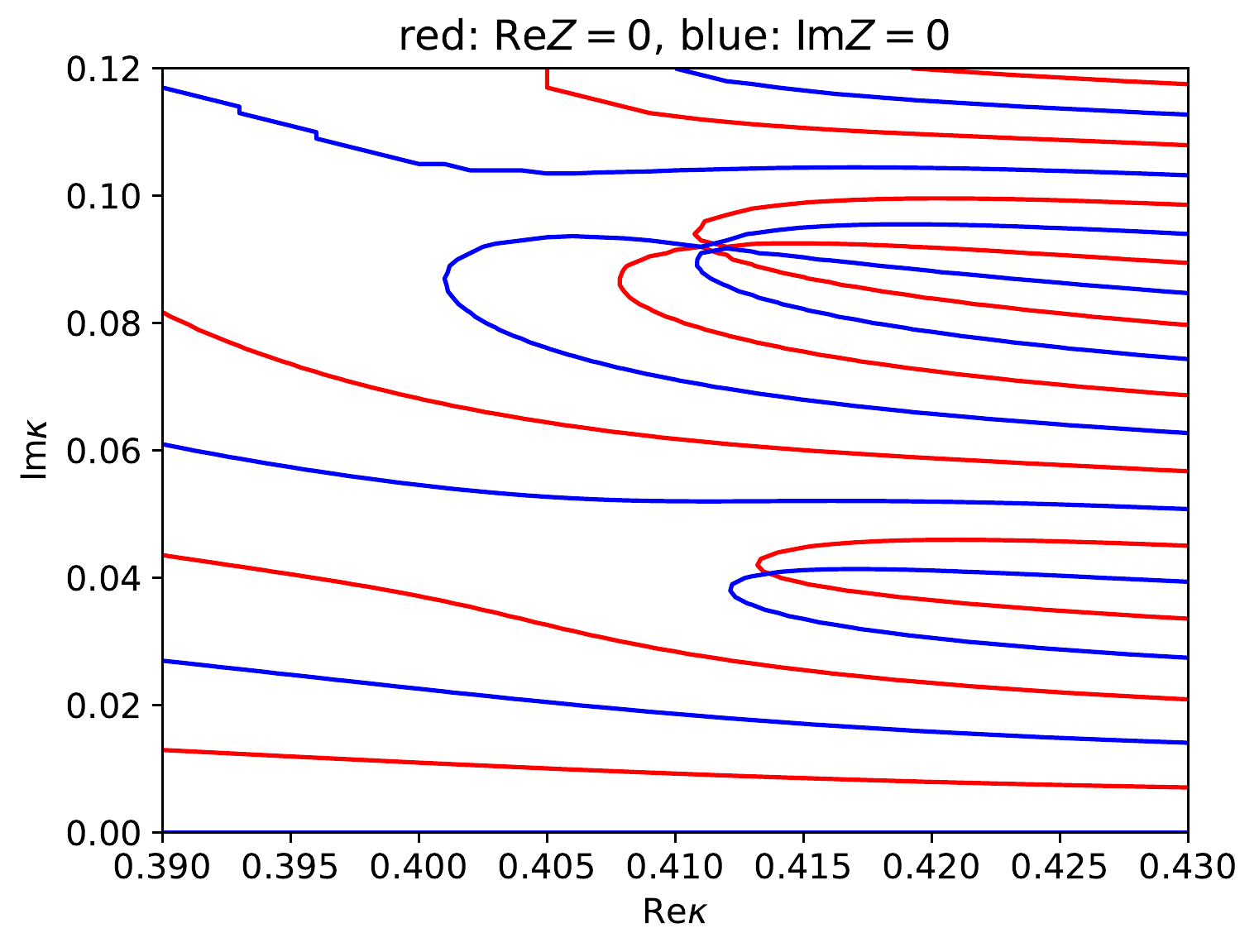}
  \caption{Partition function zeros on an $L=8$ lattice. Intersections of the red and the blue lines are the positions of $Z=0$ (Fisher zeros).}
  \label{figs:zeros_2dwilsonmajorana_iter3}
\end{figure}

\begin{figure}[htbp]
  \centering
  \includegraphics[width=0.8\hsize]{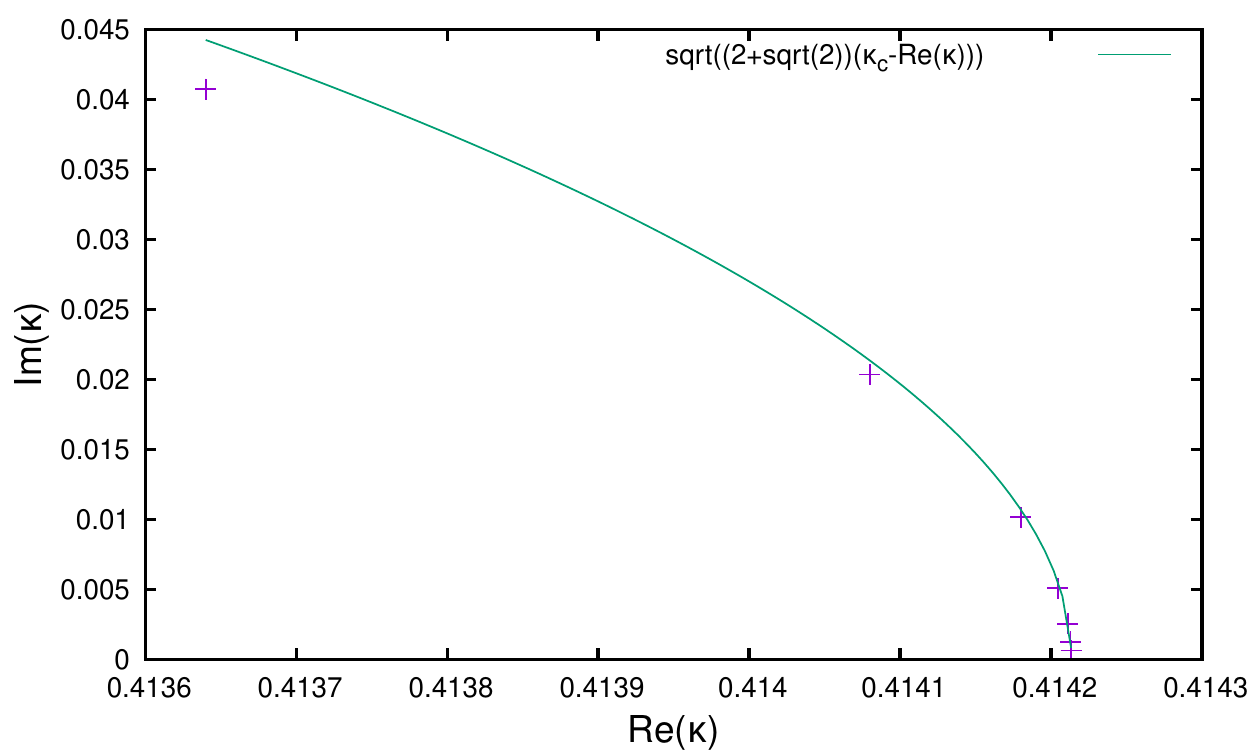}
  \caption{Fisher zeros in the complex $\kappa$ plane for $L=16$ (top),..., $L=1024$ (bottom).}
  \label{figs:firstzeros}
\end{figure}

\begin{figure}[htbp]
  \centering
  \includegraphics[width=0.8\hsize]{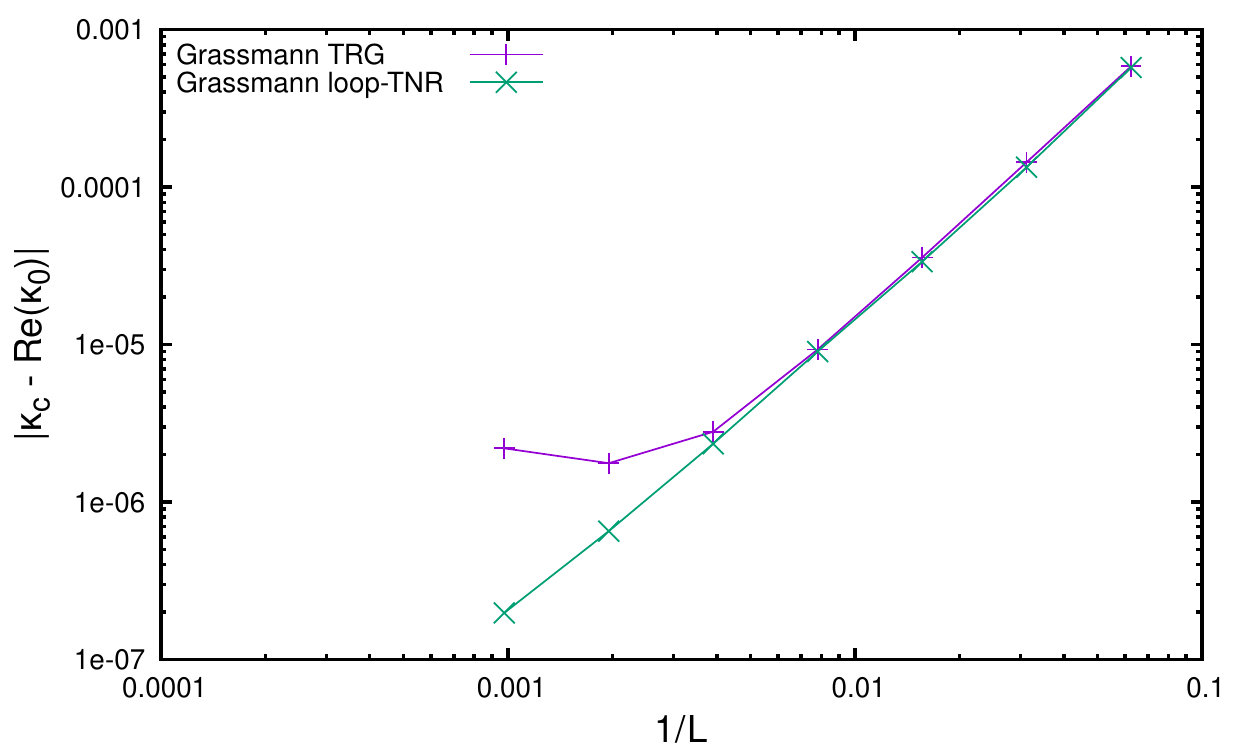}
  \caption{Difference between $\kappa_{\mathrm{c}}=\sqrt{2}-1$ and the real part of the position of the first Fisher zero.}
  \label{fig:refirstzeros_2dwilsonmajorana}
\end{figure}

\begin{figure}[htbp]
  \centering
  \includegraphics[width=0.8\hsize]{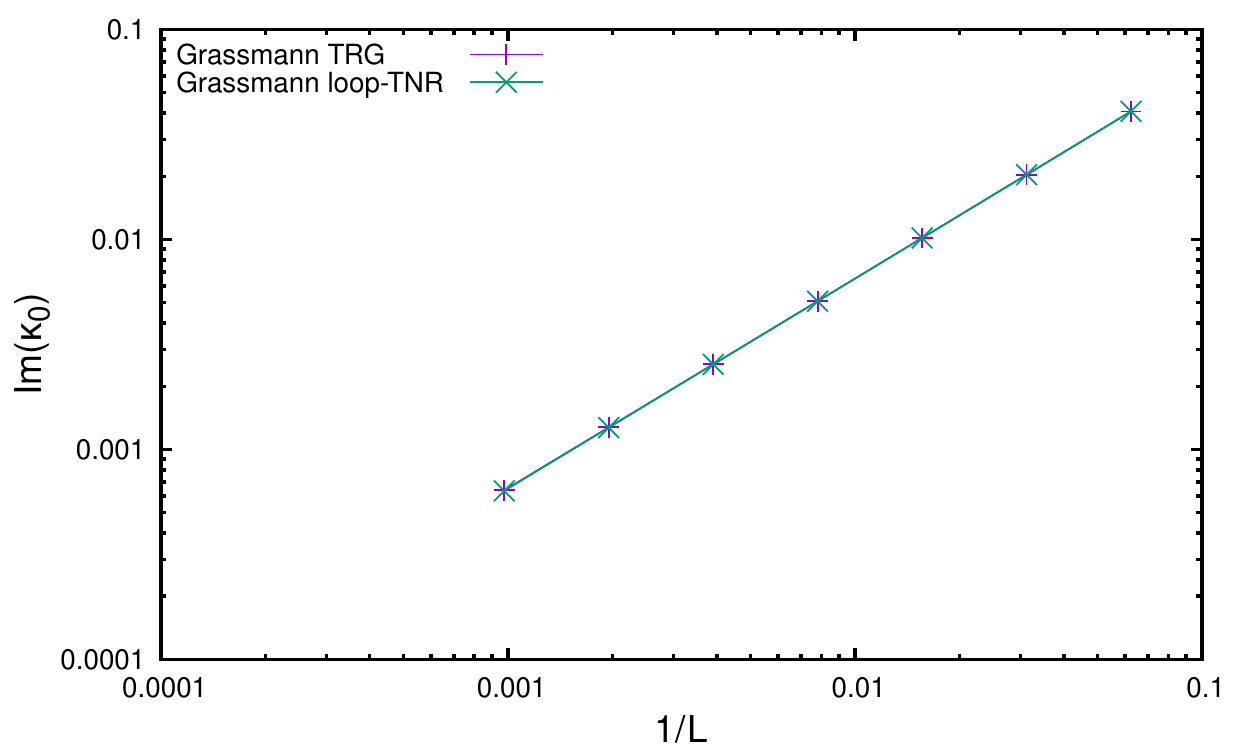}
  \caption{Imaginary part of the position of the first Fisher zero.}
  \label{fig:imfirstzeros_2dwilsonmajorana}
\end{figure}

To explore a finite size scaling behavior one can examine the Fisher zeros, the zeroes of the partition
function in the plane of complex $\kappa$~\cite{fisher1965nature}.
As an example, fig.~\ref{figs:zeros_2dwilsonmajorana_iter3} shows the zeros on an $L=8$ lattice.
In the figure, the red and the blue lines show the zeros of the real and the imaginary part of the partition function, respectively,
so that the crossing points of them can be considered to be the positions of Fisher zeros.
In the following we define the position of the first Fisher zero $\kappa_{0}$ as the coordinate of the Fisher zero which is the closest to the real axis.
In figure~\ref{figs:firstzeros} the first Fisher zeros are located in the complex $\kappa$ plane for several $L$,
where one can see that the numerical data lie on the theoretical curve $\mathrm{Im}(\kappa) = \sqrt{(2+\sqrt{2})(\kappa_{\mathrm{c}}-\mathrm{Re}(\kappa))}$ in larger volumes.
Figures~\ref{fig:refirstzeros_2dwilsonmajorana}--\ref{fig:imfirstzeros_2dwilsonmajorana} show the volume dependence of $\kappa_{0}$ in detail.
The bond dimension is set to 64 for the Grassmann TRG and up to 16 for the Grassmann loop-TNR.
In fig.~\ref{fig:refirstzeros_2dwilsonmajorana} we see that the Grassmann loop-TNR locates the critical point
$\kappa_{\mathrm{c}}=\sqrt{2}-1$ quite well while the Grassmann TRG fails as $L \rightarrow \infty$.
As for the imaginary part in fig.~\ref{fig:imfirstzeros_2dwilsonmajorana}, one cannot observe the difference between the Grassmann TRG results and the Grassmann loop-TNR ones at this resolution.
The behavior of the imaginary part is such that the first zero approaches the real
axis according to $L^{-1/\nu}$ where the observed critical exponent $\nu=1$ is consistent with the known behavior of the two dimensional Ising model.

\begin{figure}[htbp]
  \centering
  \begin{minipage}{0.32\hsize}
    \includegraphics[width=\hsize]{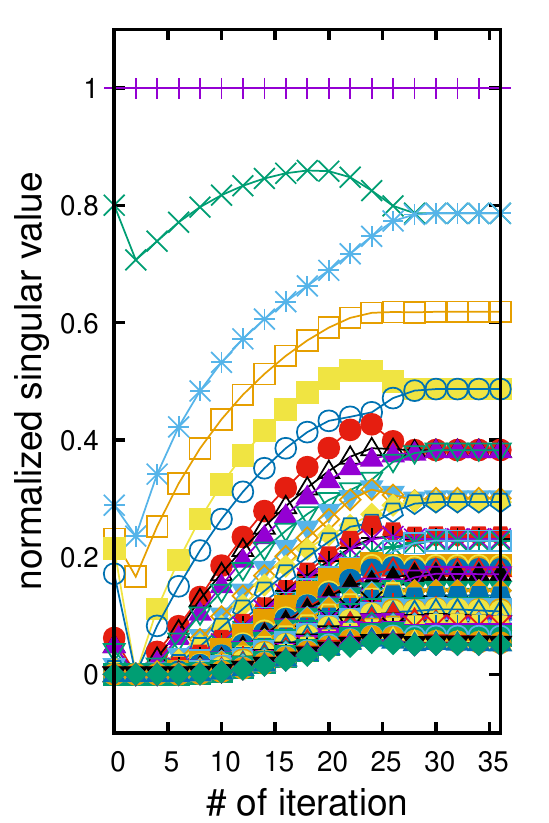}
  \end{minipage}
  \begin{minipage}{0.32\hsize}
    \includegraphics[width=\hsize]{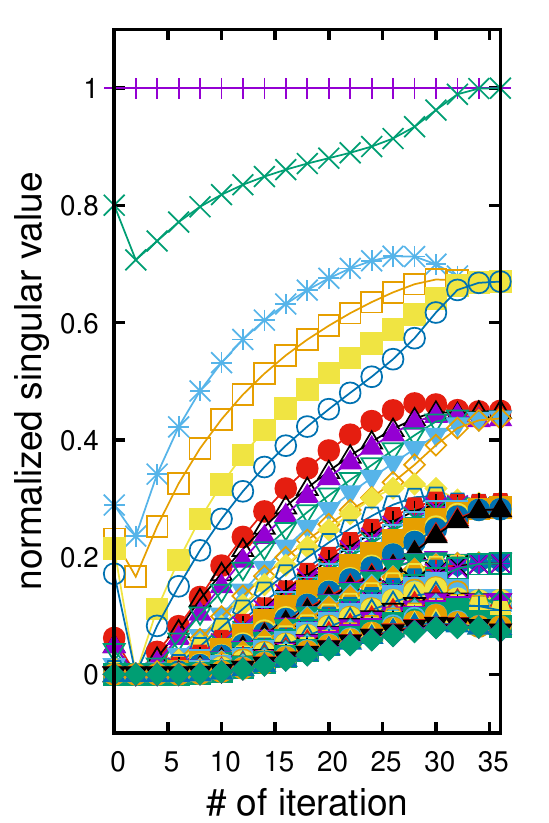}
  \end{minipage}
  \begin{minipage}{0.32\hsize}
    \includegraphics[width=\hsize]{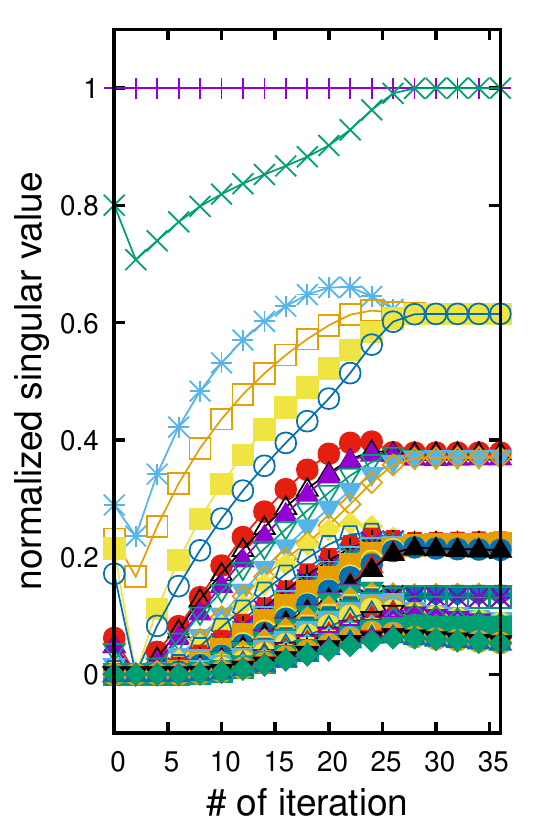}
  \end{minipage}
  \caption{Singular values obtained by Grassmann TRG at $\kappa=0.9999\kappa_{\mathrm{c}}$ (left), $\kappa=\kappa_{\mathrm{c}}$ (middle), and $\kappa=1.0001\kappa_{\mathrm{c}}$ (right). The bond dimension is 64.}
  \label{fig:svals_2dwilsonmajorana_bdim64}
\end{figure}

\begin{figure}[htbp]
  \centering
  \begin{minipage}{0.32\hsize}
    \includegraphics[width=\hsize]{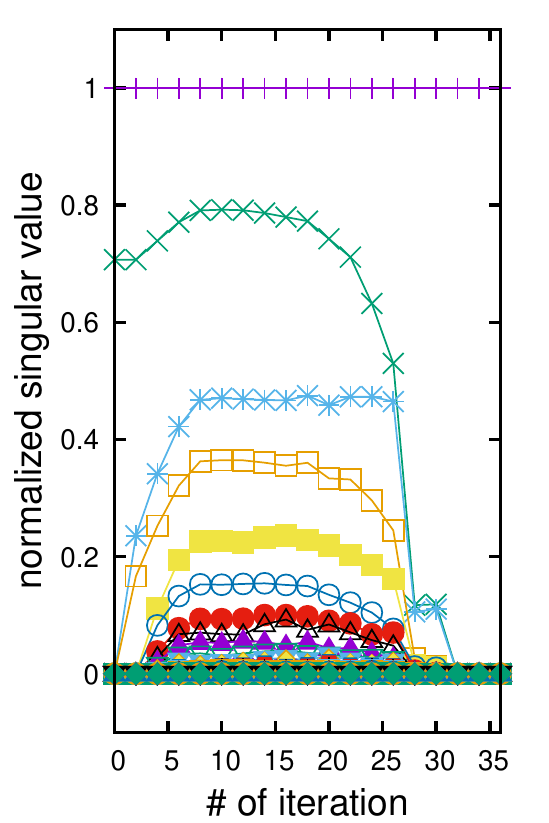}
  \end{minipage}
  \begin{minipage}{0.32\hsize}
    \includegraphics[width=\hsize]{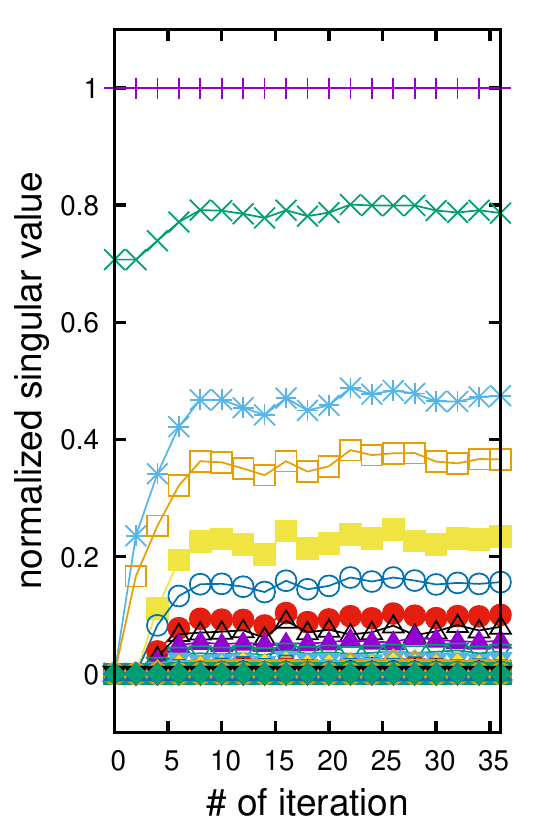}
  \end{minipage}
  \begin{minipage}{0.32\hsize}
    \includegraphics[width=\hsize]{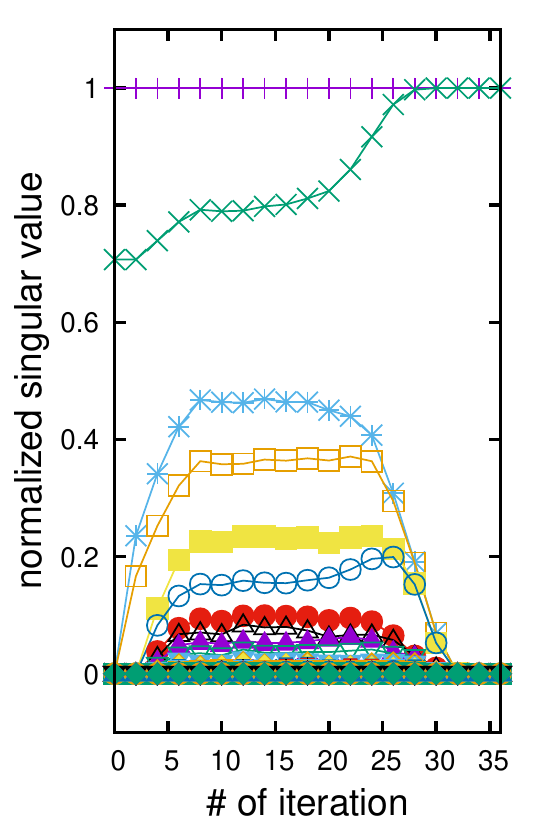}
  \end{minipage}
  \caption{Singular values obtained by Grassmann loop-TNR at $\kappa=0.9999\kappa_{\mathrm{c}}$ (left), $\kappa=\kappa_{\mathrm{c}}$ (middle), and $\kappa=1.0001\kappa_{\mathrm{c}}$ (right). The bond dimension is 16.}
  \label{fig:svals_2dwilsonmajorana_bdim16}
\end{figure}

\begin{figure}[htbp]
  \centering
  \begin{minipage}{0.32\hsize}
    \includegraphics[width=\hsize]{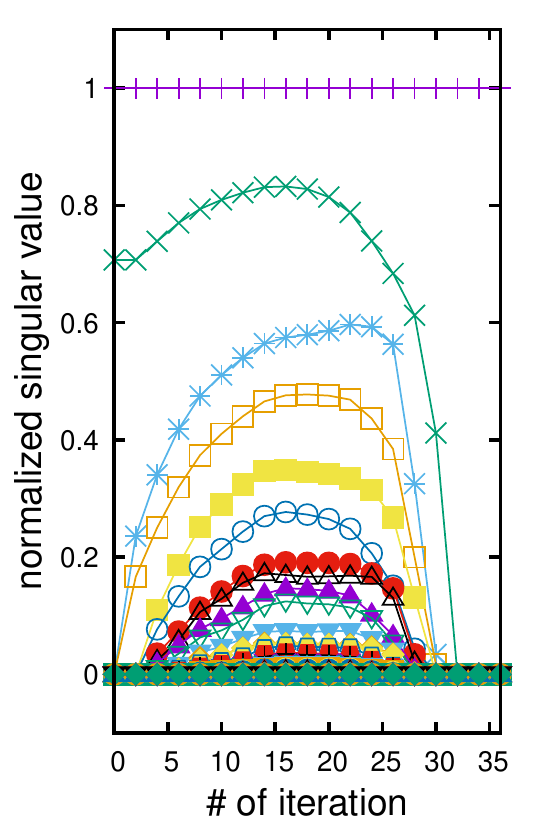}
  \end{minipage}
  \begin{minipage}{0.32\hsize}
    \includegraphics[width=\hsize]{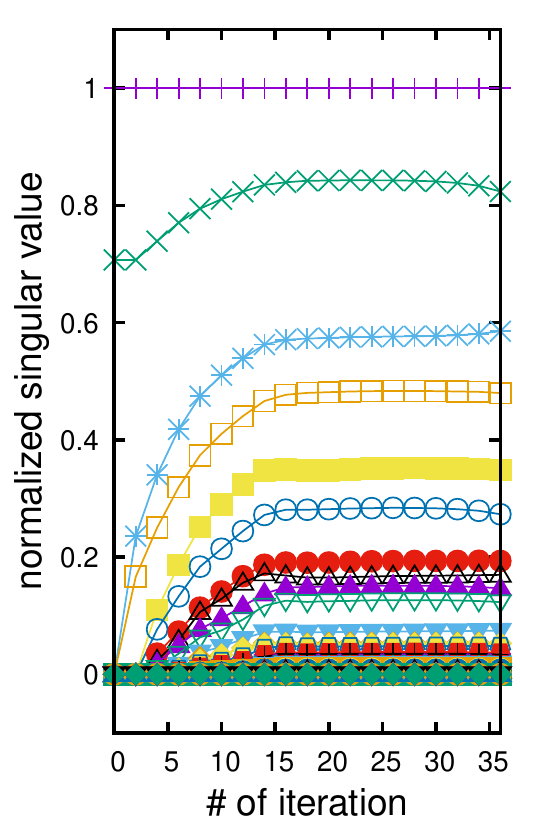}
  \end{minipage}
  \begin{minipage}{0.32\hsize}
    \includegraphics[width=\hsize]{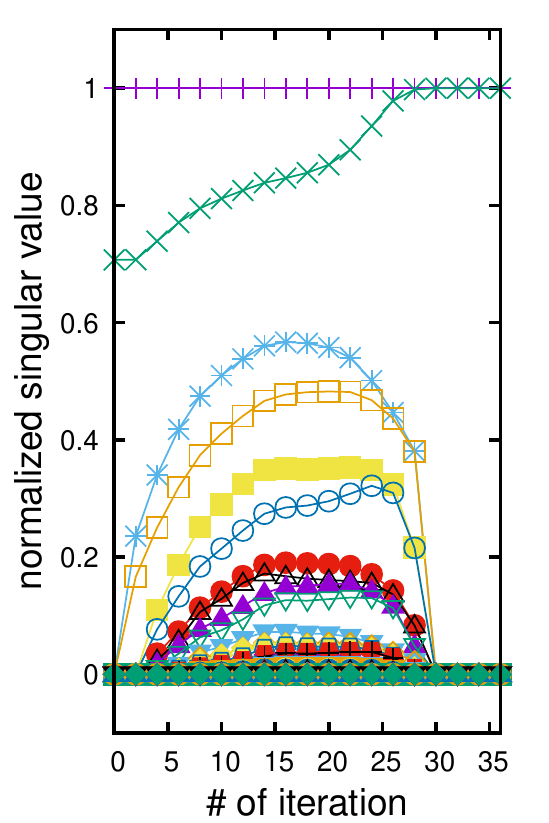}
  \end{minipage}
  \caption{Singular values obtained by Grassmann gilt-TNR at $\kappa=0.9999\kappa_{\mathrm{c}}$ (left), $\kappa=\kappa_{\mathrm{c}}$ (middle), and $\kappa=1.0001\kappa_{\mathrm{c}}$ (right). The bond dimension is 64, and $\epsilon_{\mathrm{gilt}}=10^{-6}$}
  \label{fig:svals_2dwilsonmajorana_bdim64_gilteps1e-06}
\end{figure}

The renormalization group flow of singular values obtained by the plain Grassmann TRG and the improved algorithms are shown in figs.~\ref{fig:svals_2dwilsonmajorana_bdim64}--\ref{fig:svals_2dwilsonmajorana_bdim64_gilteps1e-06}.
In the figures the singular values are normalized by the largest one at each iteration step,
and the horizontal axis is the number of iterations.
Notice that the number of iterations $i$ and the linear system size $L$ are related as $L = 2^{i+1}$~\footnote{
  Note that the number of lattice sites is reduced by $1/2$ through a single iteration step.
}.
Clearly the fixed point structure as evidenced by the behavior of the singular values
is not reproduced both on and off critical point for the vanilla Grassmann TRG algorithm.
This is due to a contamination by short range information that is not properly removed by this
algorithm.
By contrast, the improved methods show trivial fixed point structures off critical $\kappa = 0.9999\kappa_{\mathrm{c}}$ and $1.0001\kappa_{\mathrm{c}}$;
in the high temperature (symmetric) phase the number of significant singular values is one
while in the low temperature (broken) phase there are two degenerate singular values.
The fixed point structure is nontrivial only at the criticality,
where one can find a scale invariance of the hierarchical structure of the singular values that remains the same against the increasing iteration number. This fact shows that only the
improved algorithms are capable of being interpreted as implementing a true renormalization
group transformation.

\begin{figure}[htbp]
  \centering
  \includegraphics[width=0.8\hsize]{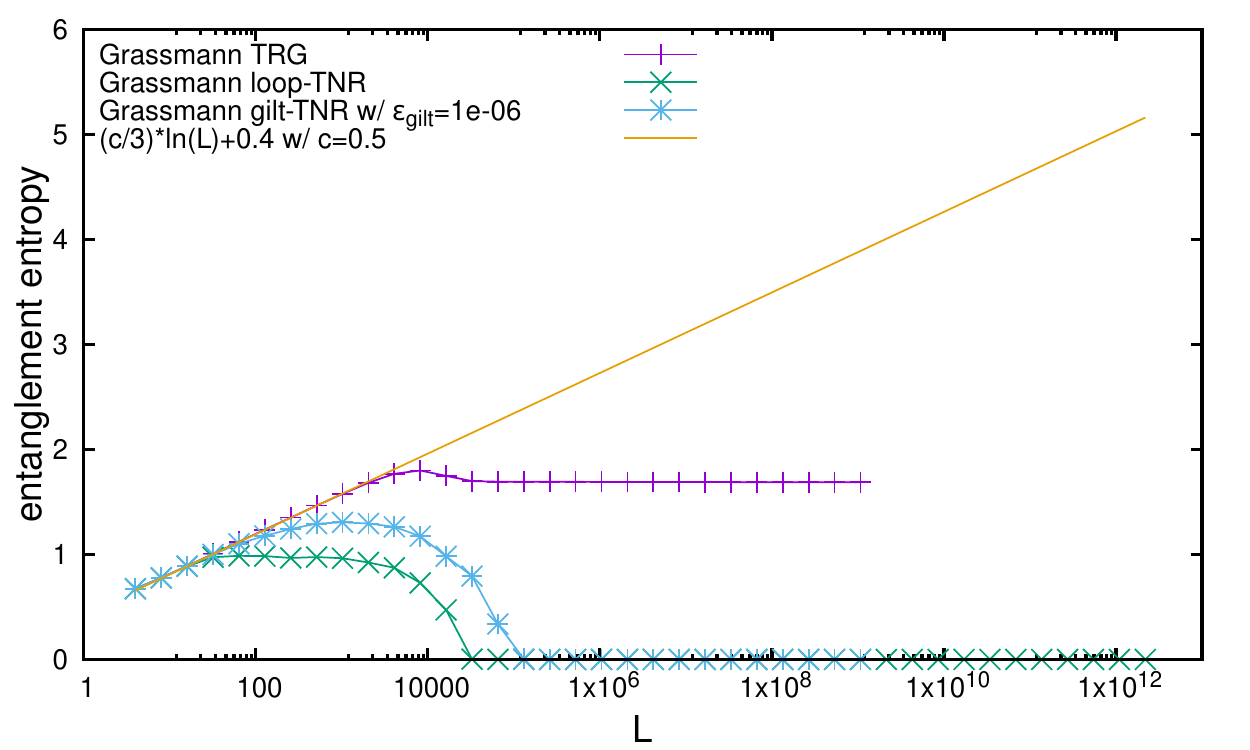}
  \caption{Entanglement entropies at $\kappa=0.9999\kappa_{\mathrm{c}}$.}
  \label{fig:ententropy_2dwilsonmajorana_kappa0.4141721410}
\end{figure}

\begin{figure}[htbp]
  \centering
  \includegraphics[width=0.8\hsize]{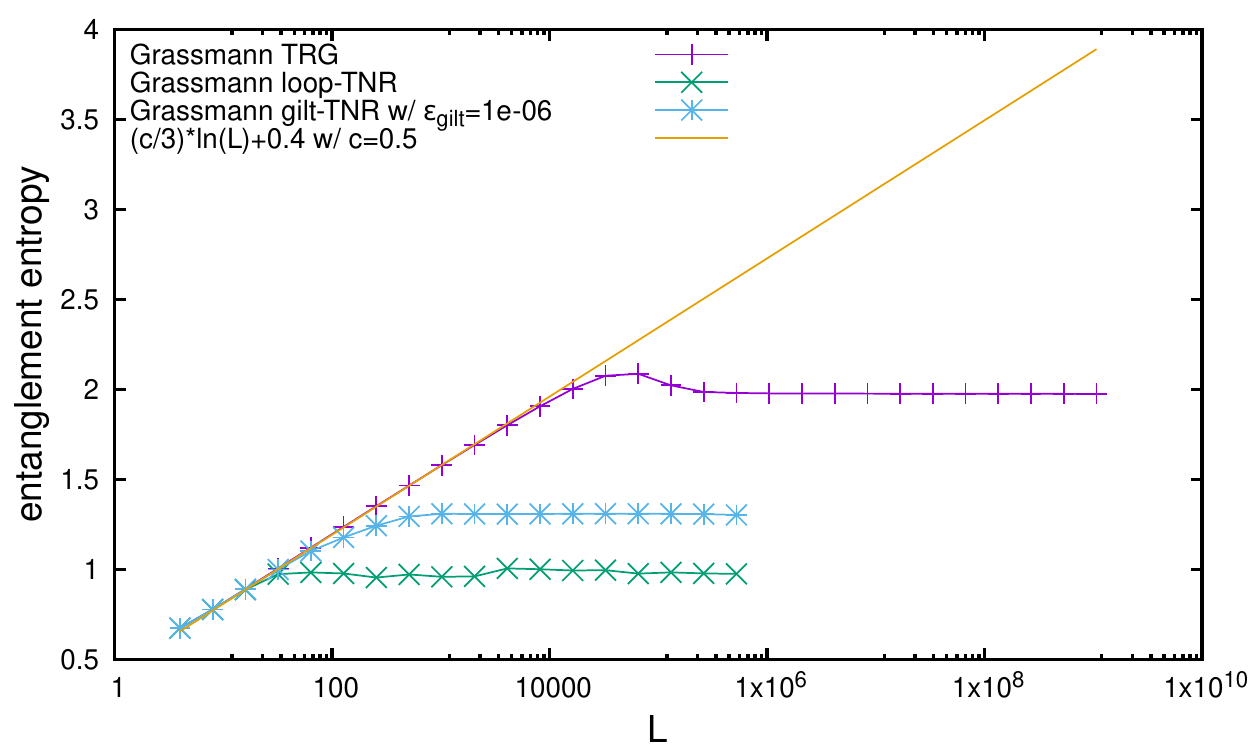}
  \caption{Entanglement entropies at $\kappa=\kappa_{\mathrm{c}}$.}
  \label{fig:ententropy_2dwilsonmajorana_kappa0.4142135624}
\end{figure}

\begin{figure}[htbp]
  \centering
  \includegraphics[width=0.8\hsize]{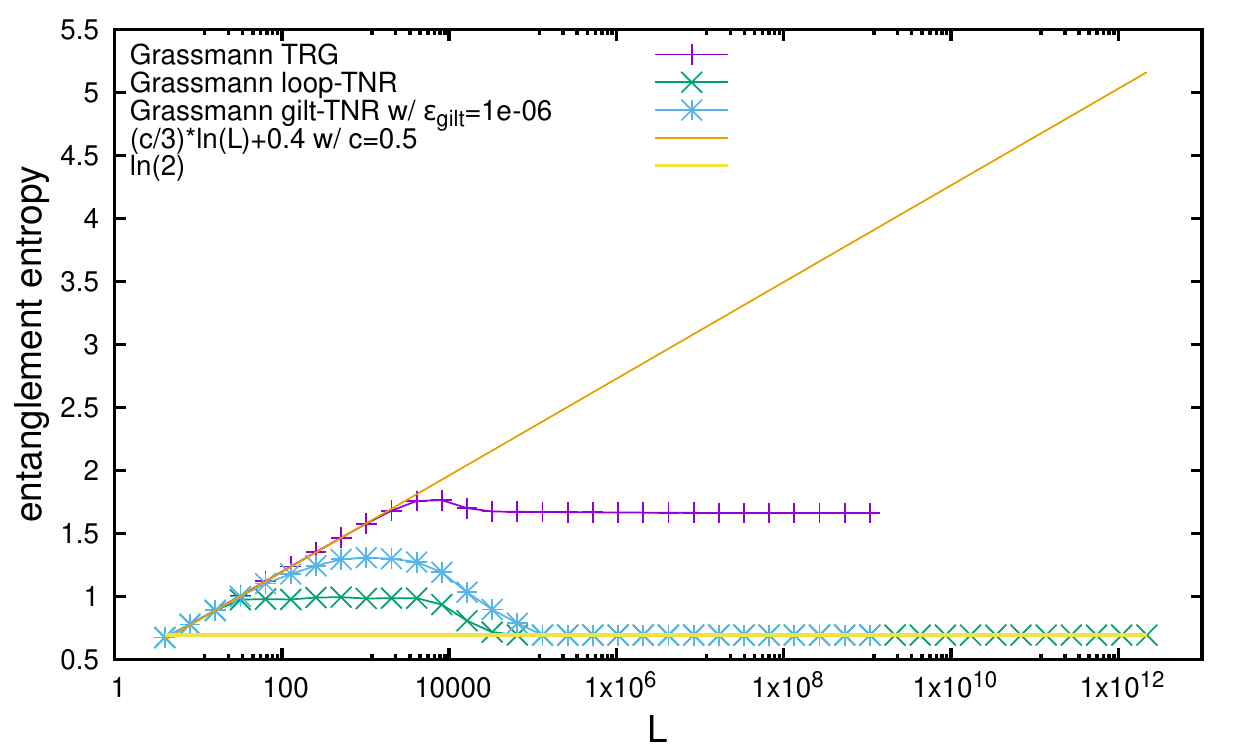}
  \caption{Entanglement entropies at $\kappa=1.0001\kappa_{\mathrm{c}}$.}
  \label{fig:ententropy_2dwilsonmajorana_kappa0.4142549837.pdf}
\end{figure}

The volume dependence of the entanglement entropy (EE) is shown for several temperatures in figs.~\ref{fig:ententropy_2dwilsonmajorana_kappa0.4141721410}--\ref{fig:ententropy_2dwilsonmajorana_kappa0.4142549837.pdf}.
To calculate the entanglement entropy, we take the eigenvalue decomposition of a reduced density matrix and use its spectra just as in the case of ref.~\cite{Yang:2015rra}.
The prediction associated with the Calabrese--Cardy formula is also shown, where $\mathrm{(EE)}=(c/3)\ln L+\text{const.}$ with the central charge $c=1/2$.
The bond dimension is 64 for the plain Grassmann TRG and the Grassmann gilt-TNR while
it is 16 for the Grassmann loop-TNR.
The Grassmann TRG goes to a nontrivial fixed point regardless of the temperature.
By contrast, the Grassmann loop-TNR and the Grassmann gilt-TNR reproduce the trivial fixed points with $\mathrm{(EE)}=0$ at the high temperature phase and $\text{(EE)}=\ln 2$ at the low temperature phase. This is consistent with theoretical expectations---the improved methods
subtract out the short range entanglement that provides a part of the entanglement
entropy at the criticality.
Thus, one can expect that the more effective the disentangling process is the lower the entropy is.

\section{Results for staggered $N_{\mathrm{f}}=2$ Gross--Neveu model}

The Wilson--Majorana fermion system is a free theory,
so this fact motivates us to assess how well the improved algorithm does in interacting fermion models.
To this end we have also applied it to the staggered $N_{\mathrm{f}}=2$ Gross--Neveu model whose action is given by
\begin{align}
  S = \sum_{n} \left\{
  \frac{1}{2} \sum_{\mu=1}^{2} \eta_{n,\mu}
  \psi^{\mathrm{T}} \left( \partial^{\mathrm{S}}_{\mu}\psi \right)_{n}
  - U\psi_{n,1}\psi_{n,2}\psi_{n,3}\psi_{n,4}
  \right\},
\end{align}
where $\psi$ is a four component staggered field and where $\eta$ is the usual
staggered fermion sign factor $\eta_{n,\mu}=\left(-1\right)^{\sum_i^{\mu-1} n_i}$.
In dimensions that are the same or greater than three, this model is capable of
generating mass without spontaneous symmetry breaking, a phenomenon which
has been addressed in several recent theoretical and numerical works~\cite{BenTov:2014eea,Ayyar:2014eua,Ayyar:2015lrd,Ayyar:2016lxq,Catterall:2015zua,Catterall:2016dzf,Ayyar:2017qii,Schaich:2017czc}.
It is believed that such symmetric mass generation may also
be important for construction of
certain lattice
chiral fermion theories in four dimensions~\cite{Catterall:2020fep,Catterall:2022jky}.
Recently a Hamiltonian formulation of this model (and for other values
of $N_{\mathrm{f}}$) has been investigated using quantum simulation and methods based on matrix product states by the authors~\cite{Asaduzzaman:2022bpi}.

The tensor network representation of the partition function in this model has the same shape with~\eqref{eq:tnrep} where the $(ijkl)$-element of the bosonic tensor at a site $n=(n_{1}, n_{2})$ is defined by
\begin{align}
  \int \mathrm{d}\psi_{1} \mathrm{d}\psi_{2} \mathrm{d}\psi_{3} \mathrm{d}\psi_{4}
  & e^{U\psi_{1}\psi_{2}\psi_{3}\psi_{4}}
    \left[ \prod_{s=1}^{4} \left( \frac{\eta_{n,1}}{2} \right)^{x_{n,s}} \right]
    \left[ \prod_{s=1}^{4} \left( \frac{\eta_{n,2}}{2} \right)^{t_{n,s}} \right] \nonumber \\
  & \cdot \psi_{4}^{l_{4}} \psi_{3}^{l_{3}} \psi_{2}^{l_{2}} \psi_{1}^{l_{1}}
    \psi_{4}^{k_{4}} \psi_{3}^{k_{3}} \psi_{2}^{k_{2}} \psi_{1}^{k_{1}}
    \psi_{4}^{j_{4}} \psi_{3}^{j_{3}} \psi_{2}^{j_{2}} \psi_{1}^{j_{1}}
    \psi_{4}^{i_{4}} \psi_{3}^{i_{3}} \psi_{2}^{i_{2}} \psi_{1}^{i_{1}}.
\end{align}
Note that, owing to the staggered sign factor $\eta_{n,2}$, the tensor network has a stripe pattern at this initial stage.
One can easily show that this pattern becomes a checkerboard and an uniform after the first and the second coarse-graining step, respectively.

\begin{figure}[htbp]
  \centering
  \includegraphics[width=0.8\hsize]{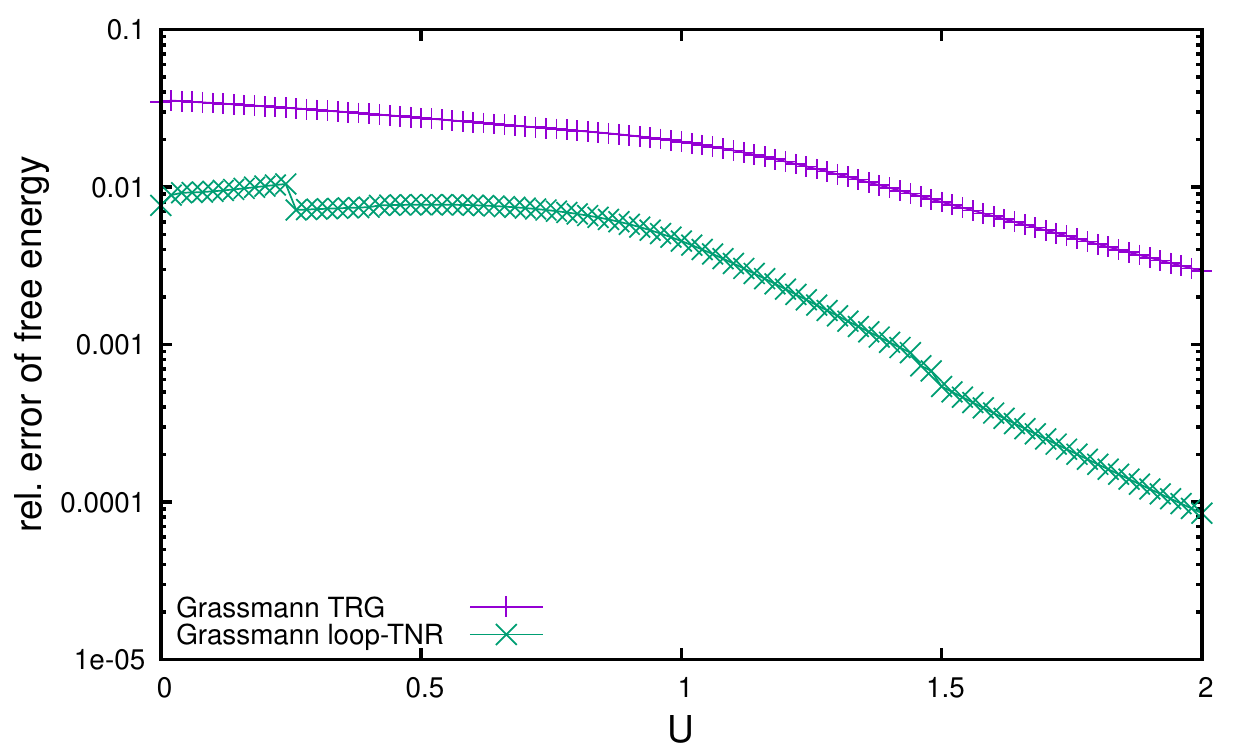}
  \caption{
    Relative errors of the free energy.
    The bond dimension is set to $64$ for both coarse-graining methods.
    The lattice volume is $L=4$, and the antiperiodic boundary conditions are imposed for both directions.
  }
  \label{fig:relerrfeneg_2dso4fermion_bdim64_iter1_apbc}
\end{figure}

Figure~\ref{fig:relerrfeneg_2dso4fermion_bdim64_iter1_apbc} shows the relative errors of the free energy calculated by the Grassmann TRG and the Grassmann loop-TNR on an $L=4$ lattice.
The analytical solutions of the model on some small lattices are shown in ref.~\cite{Ayyar:2017qii}.
With a fixed bond dimension $64$, the Grassmann loop-TNR clearly
outperforms the Grassmann TRG in the whole range of the coupling constant $U$.
This result again shows the superiority of the improved Grassmann tensor network algorithm over the vanilla Grassmann TRG method in an interacting fermion system.

\begin{figure}[htbp]
  \centering
  \includegraphics[width=0.8\hsize]{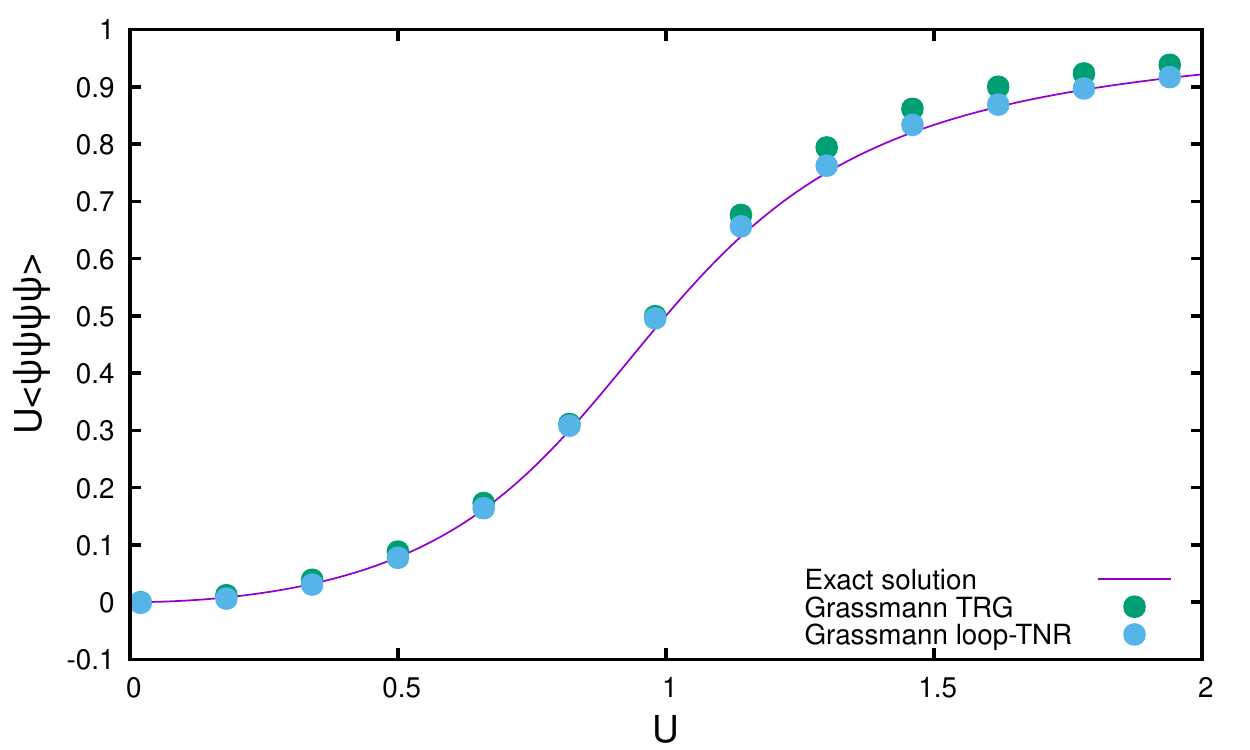}
  \caption{
    Four fermion condensate $U\left<\psi_{1}\psi_{2}\psi_{3}\psi_{4}\right>$ compared with the exact solution.
    The bond dimension is set to $64$ for both coarse-graining methods.
    The lattice volume is $L=4$, and the antiperiodic boundary conditions are imposed for both directions.
  }
  \label{fig:vertdensity_2dso4fermion_bdim64_iter1_apbc}
\end{figure}

Figure~\ref{fig:vertdensity_2dso4fermion_bdim64_iter1_apbc} shows the four fermion
condensate $U\left<\psi_{1}\psi_{2}\psi_{3}\psi_{4}\right> = (U/V) (\partial \ln Z/\partial U)$ on an $L=4$ lattice obtained using the three point numerical differentiation with $\Delta U = 0.02$.
It can be seen
that the Grassmann loop-TNR lies much closer to
the exact solution as compared to the vanilla Grassmann TRG.

\section{Conclusion}

In this paper we have studied
improved coarse-graining methods adapted to the case of
Grassmann valued tensor networks describing fermionic lattice quantum field theories.
Both the Grassmann versions of loop-TNR and gilt-TNR show
improved accuracy for quantities such as the free energy
in comparison to the normal Grassmann TRG.

Our results also exhibit quite clearly the
equivalence between Wilson--Majorana fermions and the Ising model---the critical exponents
that are obtained by finite size scaling or Fisher zero analysis match closely with those of the Ising model.
Also, the fixed point structure and the entanglement entropy of the Wilson--Majorana model obtained by the improved algorithms
reproduce the known results for the Ising model.
(See refs.~\cite{2015PhRvL.115r0405E,yang2017loop,Hauru:2017tne} for the analyses on the Ising model.)

Of course, the performance of the improved coarse-graining methods depend on the details of the model.
To check the performance in an interacting system we have also applied the Grassmann loop-TNR to the staggered $N_{\mathrm{f}}=2$ Gross--Neveu model,
and also in this case the improved algorithm outperforms the normal Grassmann TRG.
Another interesting target would be the staggered Thirring model which was studied using fermion bag methods in ref.~\cite{Ayyar:2017xmi}.
This model has a critical phase and is similar to the classical XY model.
Such cases have large numbers of internal d.o.f. and will show more complicated fixed point structure.
Indeed, precise investigations on the classical XY model have been reported with the use of the loop-TNR~\cite{2021PhRvB.104p5132U,Hong_2022}.
Thus applying the improved methods to the staggered Thirring model would be a legitimate direction to gain some new insights.

With QCD as an eventual goal it is also important
to study gauge theories. So far this has only been done for fermions in the case of
the Schwinger model~\cite{Shimizu:2014uva,Butt:2019uul}.
Although one can extract discrete d.o.f. from gauge theories with a help of \textit{e.g.} character expansions,
many of them require drastic truncation at the initial stages of construction, so that tensor network representations for gauge theories contain large errors~\cite{Bazavov:2019qih}.
Then checking how the improved methods work in such cases will not be a trivial task.

\section*{Acknowledgments}
We thank the members of the QuLAT Collaboration for valuable discussions.
RS and SC are grateful to Shailesh Chandrasekharan for sharing their Monte Carlo results for
the Gross--Neveu model.
This work was supported in part by the U.S. Department of Energy (DOE) under Award Number DE-SC0019139.
This research used resources of
the Syracuse University HTC Campus Grid and NSF award ACI-1341006
and
the National Energy Research Scientific Computing Center (NERSC), a U.S. Department of Energy Office of Science User Facility located at Lawrence Berkeley National Laboratory, operated under Contract No. DE-AC02-05CH11231 using NERSC award HEP-ERCAP0020659.

% \bibliographystyle{JHEP}
% \bibliography{references}

\begin{thebibliography}{10}

\bibitem{Levin:2006jai}
  M.~Levin and C.P.~Nave, \emph{{Tensor renormalization group approach to 2D
      classical lattice models}},
  \href{https://doi.org/10.1103/PhysRevLett.99.120601}{\emph{Phys. Rev. Lett.}
    {\bfseries 99} (2007) 120601}
  [\href{https://arxiv.org/abs/cond-mat/0611687}{{\ttfamily
      cond-mat/0611687}}].

\bibitem{2012PhRvB..86d5139X}
  Z.Y.~{Xie}, J.~{Chen}, M.P.~{Qin}, J.W.~{Zhu}, L.P.~{Yang} and T.~{Xiang},
  \emph{{Coarse-graining renormalization by higher-order singular value
      decomposition}},
  \href{https://doi.org/10.1103/PhysRevB.86.045139}{\emph{Phys. Rev.}
    {\bfseries B86} (2012) 045139}
  [\href{https://arxiv.org/abs/1201.1144}{{\ttfamily 1201.1144}}].

\bibitem{Meurice:2020pxc}
  Y.~Meurice, R.~Sakai and J.~Unmuth-Yockey, \emph{{Tensor lattice field theory
      for renormalization and quantum computing}},
  \href{https://doi.org/10.1103/RevModPhys.94.025005}{\emph{Rev. Mod. Phys.}
    {\bfseries 94} (2022) 025005}
  [\href{https://arxiv.org/abs/2010.06539}{{\ttfamily 2010.06539}}].

\bibitem{Gu:2009dr}
  Z.-C.~Gu and X.-G.~Wen, \emph{{Tensor-Entanglement-Filtering Renormalization
      Approach and Symmetry Protected Topological Order}},
  \href{https://doi.org/10.1103/PhysRevB.80.155131}{\emph{Phys. Rev.}
    {\bfseries B80} (2009) 155131}
  [\href{https://arxiv.org/abs/0903.1069}{{\ttfamily 0903.1069}}].

\bibitem{2014PhRvB..89g5116U}
  H.~{Ueda}, K.~{Okunishi} and T.~{Nishino}, \emph{{Doubling of entanglement
      spectrum in tensor renormalization group}},
  \href{https://doi.org/10.1103/PhysRevB.89.075116}{\emph{Phys. Rev.}
    {\bfseries B89} (2014) 075116}
  [\href{https://arxiv.org/abs/1306.6829}{{\ttfamily 1306.6829}}].

\bibitem{2015PhRvL.115r0405E}
  G.~{Evenbly} and G.~{Vidal}, \emph{{Tensor Network Renormalization}},
  \href{https://doi.org/10.1103/PhysRevLett.115.180405}{\emph{Phys. Rev. Lett.}
    {\bfseries 115} (2015) 180405}
  [\href{https://arxiv.org/abs/1412.0732}{{\ttfamily 1412.0732}}].

\bibitem{yang2017loop}
  S.~Yang, Z.-C.~Gu and X.-G.~Wen, \emph{{Loop optimization for tensor network
      renormalization}}, {\emph{Phys. Rev. Lett.} {\bfseries 118} (2017) 110504}
  [\href{https://arxiv.org/abs/1512.04938}{{\ttfamily 1512.04938}}].

\bibitem{Hauru:2017tne}
  M.~Hauru, C.~Delcamp and S.~Mizera, \emph{{Renormalization of tensor networks
      using graph independent local truncations}},
  \href{https://doi.org/10.1103/PhysRevB.97.045111}{\emph{Phys. Rev.}
    {\bfseries B97} (2018) 045111}
  [\href{https://arxiv.org/abs/1709.07460}{{\ttfamily 1709.07460}}].

\bibitem{Wolff:2020oky}
  U.~Wolff, \emph{{Ising model as Wilson--Majorana Fermions}},
  \href{https://doi.org/10.1016/j.nuclphysb.2020.115061}{\emph{Nucl. Phys.}
    {\bfseries B955} (2020) 115061}
  [\href{https://arxiv.org/abs/2003.01579}{{\ttfamily 2003.01579}}].

\bibitem{eckart1936approximation}
  C.~Eckart and G.~Young, \emph{{The approximation of one matrix by another of
      lower rank}}, {\emph{Psychometrika} {\bfseries 1} (1936) 211}.

\bibitem{Hong_2022}
  S.~{Hong} and D.-H.~{Kim}, \emph{{Tensor network calculation of the logarithmic
      correction exponent in the XY model}},
  \href{https://doi.org/10.7566/jpsj.91.084003}{\emph{J. Phys. Soc. Jpn.}
    {\bfseries 91} (2022) 084003}
  [\href{https://arxiv.org/abs/2205.02773}{{\ttfamily 2205.02773}}].

\bibitem{Delcamp:2020hzo}
  C.~Delcamp and A.~Tilloy, \emph{{Computing the renormalization group flow of
      two-dimensional $\phi^4$ theory with tensor networks}},
  \href{https://doi.org/10.1103/PhysRevResearch.2.033278}{\emph{Phys. Rev.
      Res.} {\bfseries 2} (2020) 033278}
  [\href{https://arxiv.org/abs/2003.12993}{{\ttfamily 2003.12993}}].

\bibitem{Gu:2010yh}
  Z.-C.~Gu, F.~Verstraete and X.-G.~Wen, \emph{{Grassmann tensor network states
      and its renormalization for strongly correlated fermionic and bosonic
      states}},  \href{https://arxiv.org/abs/1004.2563}{{\ttfamily 1004.2563}}.

\bibitem{Gu:2013gba}
  Z.-C.~Gu, \emph{{Efficient simulation of Grassmann tensor product states}},
  \href{https://doi.org/10.1103/PhysRevB.88.115139}{\emph{Phys. Rev.}
    {\bfseries B88} (2013) 115139}
  [\href{https://arxiv.org/abs/1109.4470}{{\ttfamily 1109.4470}}].

\bibitem{fisher1965nature}
  M.E.~Fisher, \emph{The nature of critical points}, University of Colorado Press
  (1965).

\bibitem{Yang:2015rra}
  L.-P.~Yang, Y.~Liu, H.~Zou, Z.Y.~Xie and Y.~Meurice, \emph{{Fine structure of
      the entanglement entropy in the O(2) model}},
  \href{https://doi.org/10.1103/PhysRevE.93.012138}{\emph{Phys. Rev.}
    {\bfseries E93} (2016) 012138}
  [\href{https://arxiv.org/abs/1507.01471}{{\ttfamily 1507.01471}}].

\bibitem{BenTov:2014eea}
  Y.~BenTov, \emph{{Fermion masses without symmetry breaking in two spacetime
      dimensions}}, \href{https://doi.org/10.1007/JHEP07(2015)034}{\emph{JHEP}
    {\bfseries 07} (2015) 034} [\href{https://arxiv.org/abs/1412.0154}{{\ttfamily
      1412.0154}}].

\bibitem{Ayyar:2014eua}
  V.~Ayyar and S.~Chandrasekharan, \emph{{Massive fermions without fermion
      bilinear condensates}},
  \href{https://doi.org/10.1103/PhysRevD.91.065035}{\emph{Phys. Rev.}
    {\bfseries D91} (2015) 065035}
  [\href{https://arxiv.org/abs/1410.6474}{{\ttfamily 1410.6474}}].

\bibitem{Ayyar:2015lrd}
  V.~Ayyar and S.~Chandrasekharan, \emph{{Origin of fermion masses without
      spontaneous symmetry breaking}},
  \href{https://doi.org/10.1103/PhysRevD.93.081701}{\emph{Phys. Rev.}
    {\bfseries D93} (2016) 081701}
  [\href{https://arxiv.org/abs/1511.09071}{{\ttfamily 1511.09071}}].

\bibitem{Ayyar:2016lxq}
  V.~Ayyar and S.~Chandrasekharan, \emph{{Fermion masses through four-fermion
      condensates}}, \href{https://doi.org/10.1007/JHEP10(2016)058}{\emph{JHEP}
    {\bfseries 10} (2016) 058}
  [\href{https://arxiv.org/abs/1606.06312}{{\ttfamily 1606.06312}}].

\bibitem{Catterall:2015zua}
  S.~Catterall, \emph{{Fermion mass without symmetry breaking}},
  \href{https://doi.org/10.1007/JHEP01(2016)121}{\emph{JHEP} {\bfseries 01}
    (2016) 121} [\href{https://arxiv.org/abs/1510.04153}{{\ttfamily
      1510.04153}}].

\bibitem{Catterall:2016dzf}
  S.~Catterall and D.~Schaich, \emph{{Novel phases in strongly coupled
      four-fermion theories}},
  \href{https://doi.org/10.1103/PhysRevD.96.034506}{\emph{Phys. Rev.}
    {\bfseries D96} (2017) 034506}
  [\href{https://arxiv.org/abs/1609.08541}{{\ttfamily 1609.08541}}].

\bibitem{Ayyar:2017qii}
  V.~Ayyar and S.~Chandrasekharan, \emph{{Generating a nonperturbative mass gap
      using Feynman diagrams in an asymptotically free theory}},
  \href{https://doi.org/10.1103/PhysRevD.96.114506}{\emph{Phys. Rev.}
    {\bfseries D96} (2017) 114506}
  [\href{https://arxiv.org/abs/1709.06048}{{\ttfamily 1709.06048}}].

\bibitem{Schaich:2017czc}
  D.~Schaich and S.~Catterall, \emph{{Phases of a strongly coupled four-fermion
      theory}}, \href{https://doi.org/10.1051/epjconf/201817503004}{\emph{EPJ Web
      Conf.} {\bfseries 175} (2018) 03004}
  [\href{https://arxiv.org/abs/1710.08137}{{\ttfamily 1710.08137}}].

\bibitem{Catterall:2020fep}
  S.~Catterall, \emph{{Chiral lattice fermions from staggered fields}},
  \href{https://doi.org/10.1103/PhysRevD.104.014503}{\emph{Phys. Rev.}
    {\bfseries D104} (2021) 014503}
  [\href{https://arxiv.org/abs/2010.02290}{{\ttfamily 2010.02290}}].

\bibitem{Catterall:2022jky}
  S.~Catterall, \emph{{'t Hooft anomalies for staggered fermions}},
  \href{https://arxiv.org/abs/2209.03828}{{\ttfamily 2209.03828}}.

\bibitem{Asaduzzaman:2022bpi}
  M.~Asaduzzaman, S.~Catterall, G.C.~Toga, Y.~Meurice and R.~Sakai,
  \emph{{Quantum Simulation of the N flavor Gross--Neveu Model}},
  \href{https://arxiv.org/abs/2208.05906}{{\ttfamily 2208.05906}}.

\bibitem{Ayyar:2017xmi}
  V.~Ayyar, S.~Chandrasekharan and J.~Rantaharju, \emph{{Benchmark results in the
      2D lattice Thirring model with a chemical potential}},
  \href{https://doi.org/10.1103/PhysRevD.97.054501}{\emph{Phys. Rev.}
    {\bfseries D97} (2018) 054501}
  [\href{https://arxiv.org/abs/1711.07898}{{\ttfamily 1711.07898}}].

\bibitem{2021PhRvB.104p5132U}
  A.~{Ueda} and M.~{Oshikawa}, \emph{{Resolving the
      Berezinskii--Kosterlitz--Thouless transition in the two-dimensional XY model
      with tensor-network-based level spectroscopy}},
  \href{https://doi.org/10.1103/PhysRevB.104.165132}{\emph{Phys. Rev.}
    {\bfseries B104} (2021) 165132}
  [\href{https://arxiv.org/abs/2105.11460}{{\ttfamily 2105.11460}}].

\bibitem{Shimizu:2014uva}
  Y.~Shimizu and Y.~Kuramashi, \emph{{Grassmann tensor renormalization group
      approach to one-flavor lattice Schwinger model}},
  \href{https://doi.org/10.1103/PhysRevD.90.014508}{\emph{Phys. Rev.}
    {\bfseries D90} (2014) 014508}
  [\href{https://arxiv.org/abs/1403.0642}{{\ttfamily 1403.0642}}].

\bibitem{Butt:2019uul}
  N.~Butt, S.~Catterall, Y.~Meurice, R.~Sakai and J.~Unmuth-Yockey, \emph{{Tensor
      network formulation of the massless Schwinger model with staggered
      fermions}}, \href{https://doi.org/10.1103/PhysRevD.101.094509}{\emph{Phys.
      Rev.} {\bfseries D101} (2020) 094509}
  [\href{https://arxiv.org/abs/1911.01285}{{\ttfamily 1911.01285}}].

\bibitem{Bazavov:2019qih}
  A.~Bazavov, S.~Catterall, R.G.~Jha and J.~Unmuth-Yockey, \emph{{Tensor
      renormalization group study of the non-Abelian Higgs model in two
      dimensions}}, \href{https://doi.org/10.1103/PhysRevD.99.114507}{\emph{Phys.
      Rev.} {\bfseries D99} (2019) 114507}
  [\href{https://arxiv.org/abs/1901.11443}{{\ttfamily 1901.11443}}].

\end{thebibliography}

\providecommand{\href}[2]{#2}\begingroup\raggedright\endgroup

\end{document}